%% file: main.tex
\documentclass[sigconf]{acmart}

\AtBeginDocument{%
  \providecommand\BibTeX{{%
    \normalfont B\kern-0.5em{\scshape i\kern-0.25em b}\kern-0.8em\TeX}}}

\copyrightyear{2022}
\acmYear{2022}
\setcopyright{acmlicensed}\acmConference[ACSAC '22]{Annual Computer Security Applications Conference}{December 5--9, 2022}{Austin, TX, USA}
\acmBooktitle{Annual Computer Security Applications Conference (ACSAC '22), December 5--9, 2022, Austin, TX, USA}
\acmPrice{15.00}
\acmDOI{10.1145/3564625.3568000}
\acmISBN{978-1-4503-9759-9/22/12}

\usepackage{amsmath}
\usepackage{caption}
\usepackage{algorithmic}
\usepackage{array}
\usepackage{mdwtab}
\usepackage{eqparbox}
\usepackage{subfig}
\usepackage{xurl}
\usepackage{comment}
\usepackage[utf8]{inputenc}
\usepackage[OT1]{fontenc}
\usepackage{graphicx}
\usepackage{rotating}
\usepackage{multirow}
\usepackage{makecell}
\usepackage{xspace}
\usepackage{tabto}
\usepackage{pifont}
\usepackage{tikz}
\usepackage{booktabs}
\usetikzlibrary{shapes,arrows}
\usepackage{xcolor}
\usetikzlibrary{shapes,arrows}

\newcommand{\ie}{i.\@\,e.\@\xspace}
\newcommand{\eg}{e.\@\,g.\@\xspace}

\newcommand{\etc}{etc.\@\xspace}
\newcommand{\etal}{et~al.\@\xspace}
\newcommand{\cmark}{\ding{52}}
\newcommand{\xmark}{\ding{56}}

\newcommand{\eat}[1]{{}}

\begin{document}

\title[You have been warned: Abusing 5G's Warning and Emergency Systems]{You have been warned: \\Abusing 5G's Warning and Emergency Systems}

\author{Evangelos Bitsikas}
\affiliation{%
  \institution{New York University Abu Dhabi, Northeastern University}
  \city{Abu Dhabi, UAE \& Boston, US}
  \country{}
}
\email{evangelos.bitsikas@nyu.edu}

\author{Christina P{\"o}pper}
\affiliation{%
  \institution{New York University Abu Dhabi}
  \city{Abu Dhabi}
  \country{UAE}}
\email{christina.poepper@nyu.edu}

%% Keywords. The author(s) should pick words that accurately describe
%% the work being presented. Separate the keywords with commas.
\keywords{5G, Public Warning System, spoofing, suppression, MitM attacks}

\begin{abstract}
The Public Warning System (PWS) is an essential part of cellular networks and a country's civil protection. Warnings can notify users of hazardous events (\eg, floods, earthquakes) and crucial national matters that require immediate attention. PWS attacks disseminating fake warnings or concealing precarious events can have a serious impact, causing fraud, panic, physical harm, or unrest to users within an affected area.   
In this work, we conduct the first comprehensive investigation of PWS security in 5G networks. We demonstrate five practical attacks that may impact the security of 5G-based Commercial Mobile Alert System (CMAS) as well as Earthquake and Tsunami Warning System (ETWS) alerts. Additional to identifying the vulnerabilities, we investigate two PWS spoofing and three PWS suppression attacks, with or without a man-in-the-middle (MitM) attacker. We discover that MitM-based attacks have more severe impact than their non-MitM counterparts. Our PWS barring attack is an effective technique to eliminate legitimate warning messages. We perform a rigorous analysis of the roaming aspect of the PWS, incl.~its potentially secure version, and report the implications of our attacks on other emergency features (\eg, $911$ SIP calls). We discuss possible countermeasures and note that eradicating the attacks necessitates a scrupulous reevaluation of the PWS design and a secure implementation.
\end{abstract}

\maketitle

\input{sections/1.introduction}
\input{sections/2.background}
\input{sections/3.adversarial-setup}
\input{sections/4.exploitation}
\input{sections/5.experimentation}
\input{sections/6.countermeasures}
\input{sections/8.related-work}
\input{sections/9.conclusion}

\begin{acks}

This work was supported by the Abu Dhabi Award for Research Excellence (AARE) 2019 (\#AARE19-236) and the Center for Cyber Security at New York University Abu Dhabi (NYUAD).

\end{acks}

\bibliographystyle{ACM-Reference-Format}
\bibliography{main.bib}

\appendix

\input{sections/appendix}

\end{document}

%% file: sections/1.introduction.tex
\section{Introduction}

An integral part of cellular networks is the Public Warning System (PWS) which is responsible for alerting users about emergencies and hazardous events. Each country has its own PWS as a critical component of national and civil security. In the US, the Federal Communication Commission (FCC) and the Federal Emergency Management Agency (FEMA) of the Homeland Security have explored ways to enhance their alerting capabilities~\cite{federal_communications_commission_2021, homeland_security_practices, NAP24935}, while the EU has launched and keeps improving its own PWS based on the European Telecommunications Standards Institute (ETSI) and European Commission directives~\cite{eu_commission, European_assoc, 3gpp.102.900}. In addition, ETSI and the Third Generation Partnership Project (3GPP) have been working on the specifications of the enhanced Public Warning System (ePWS)~\cite{3gpp.122.268} that is supposed to be compatible with all prior generations and to improve the comprehension of warning notifications. We provide more information on the adoption of the PWS in App.~\ref{pws-adopt}.

In general, the warning system utilizes paging messages to force a User Equipment (UE) to enter the RRC-Connected state (in case the UE is in Idle or Inactive state). Once active, it can receive warning messages that are transmitted via System Information Block (SIB) messages. The SIB messages belong to the broadcast transmissions that are mainly used to facilitate the initial connection to the Radio Access Network (RAN) and Core Network, and to assist the mobility and critical operations. PWS' principal way to notify the user is through these cell broadcasts, as alternative (\eg, SMS-based) ways are currently not prevalent and utilize different network procedures (\eg, are service-based).

Current PWS deployments realize the 3GPP standards but lack security properties such as warning verification and integrity protection~\cite{3gpp.123.041, 3gpp.138.331, 3gpp.133.969}; the associated security flaws began on the early design of the PWS for legacy generations and remain broadly unresolved to date. In particular, paging and SIB messages are unprotected. False alarms were reported~\cite{reuters20:ballistic-alert, CTV20:nuclear-alert} and spam attacks for profit~\cite {CNN21:Philippines-alerts} unveiled concerns about the stability and effectiveness of the current PWS. Generally, by spoofing and tampering with warning messages an adversary can spread panic among a population in a targeted area to stimulate terrorist activities, impede civil protection by security agencies, and profit through spam or fraud. Not less perilous is the suppression of legitimate warning messages, which would thwart message reception about an emergency incident for users (\eg, natural disasters~\cite{owid14:natural-disasters}). It is equivocal if security enhancements are going to supplement the PWS in the future. 

A first academic notice of weaknesses of the public warning system was made by Lee for LTE \etal~\cite{Lee19:spoofing-alerts-lte} where the authors evaluated the spoofing of Commercial Mobile Alert System (CMAS) messages while focusing exclusively on LTE's Presidential Alerts. \cite{Lee19:spoofing-alerts-lte} provides insights into exemplary existing weaknesses, but does not systematically explore the full potential of the attacker (\eg, there is no consideration of a MitM attacker). 

In our paper, we demonstrate that the impact of attacks on PWS is significantly higher than portrayed so far. In particular, we discovered that a MitM attacker can have a large attacking window, not being limited to $42$ sec (with warning periodicity equal to $160$ msec) and $262$ transmissions as described in~\cite{Lee19:spoofing-alerts-lte}. In fact, an attacker that successfully exploits the cell reselection and handover procedures to set up a MitM can inflict further damage regardless of the Authentication and Key Agreement procedure (AKA) by extending the attacking period and hence the number of spoofed alerts. Additionally, we introduce and investigate an attack we call \emph{PWS Barring Attack} that can efficiently cause warning suppression without the requirements of a redundant malicious attachment to a fake base station. The PWS barring attack can be used as a less intrusive and convenient Denial-of-Service (DoS) attack by an attacker.

Prior works (5GReasoner~\cite{Hussain19:5G-reasoner}, LTEInspector~\cite{Hussain18:LTE-Inspector}, Touching the Untouchables~\cite{Kim19:touching-untouchables}) have investigated RRC and NAS vulnerabilities, but are limited to open-source tools for LTE (\eg, srsLTE) and theoretical protocol evaluations for 5G. Practical 5G experimentation on PWS vulnerabilities and attacks have not been conducted. We thus focus on 5G Standalone (SA) and non-Standalone (NSA) systems, and experimentally validate vulnerabilities and attacks -- new and old -- considering the latest defenses, updates, and specifications. Our experiments with a 5G SA setup (Amarisoft-box) provide reliable results from commercial HW/SW. We present a thorough study of the PWS exploring two attacker setups and multiple attacks (spoofing \& suppression variations). In our experiments, we utilize all types of warning messages and observe the behaviour of $5$ smartphone devices from different manufacturers to assess the impact of each attack. Due to the importance of roaming for users located outside of their home network, we additionally analyze PWS security in conjunction with roaming and delve into countermeasures to mitigate the identified vulnerabilities. 

In short, our major contributions are as follows:
\begin{enumerate}
    \item We investigate the security of the 5G PWS system considering the latest defenses, updates and 5G specifications, test all warning types defined by the 3GPP and used in real-life PWS, Earthquake and Tsunami Warning System (ETWS) messages and CMAS messages. Through 5G SA experimentation, we comprehensively list involved vulnerabilities and security deficiencies that allow an attacker to exploit the 5G PWS.
    \item We explore multiple attack vectors in depth: \textit{(a)} We perform \emph{PWS spoofing} and \emph{PWS suppression} attacks based on two different setups, MitM based and non-MitM based. We reveal that when the attacker adopts the MitM deployment, the impact is larger, meaning the spoofing window is greater than in non-MitM situations. \textit{(b)} We present the \emph{PWS Barring Attack} that can be used for effective warning suppression. We discover that it is characterized by a greater impact and feasibility than other suppression attacks.
    \item We thoroughly analyze the combination of warning messages with the roaming feature of cellular networks. Given possible countermeasures against PWS attacks, we examine the effects of our attacks on the current roaming deployment and potentially secure version of the PWS. 
\end{enumerate}

We also assess the impact of our attacks on the user including effects on the SMS-based warnings and emergency calls. We provide an extensive list of possible countermeasures while pointing out advantages and drawbacks when implemented in the PWS.

\textbf{Responsible Disclosure.} Due to the significance of the emergency systems and their broad implications, we reported our findings to GSMA, the GSM Association (disclosure date: Feb 7, 2022). GSMA has acknowledged them under the number CVD-2022-0054, separately notified 3GPP, and has issued an associated briefing paper to share with its members. We were in active exchange with GSMA for clarifications and brainstorming about countermeasures, and we also disclosed our results to several agencies (CISA, FEMA, ENISA) so they can inform affected organizations about the dangers and defensive actions.

%% file: sections/2.background.tex
\section{Background}

In this section, we summarize the structure and functionality of the PWS on 5G network systems according to the specifications~\cite{3gpp.123.041, 3gpp.138.331}.

\subsection{Network Structure}

\begin{figure}[bt]
    \centering
    \includegraphics[width=1.\columnwidth,keepaspectratio]{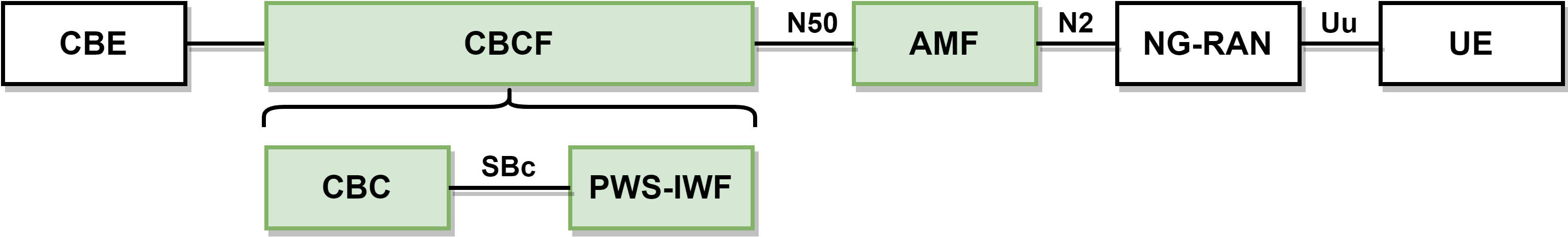}
    \caption{5G PWS Architecture}
    \label{fig:pwsnetwork}
\end{figure}

The network architecture is presented in Figure~\ref{fig:pwsnetwork}. It consists of the following entities and functions: 

\noindent\textbf{CBE} (Cell Broadcast Entity). The CBE's responsibility is to properly format the Cell Broadcast Service (CBS) messages and when necessary divide the CBS message into a number of pages. A federal authority typically informs the CBE about the corresponding warning message. 

\noindent\textbf{CBC/CBCF} (Cell Broadcast Center/Cell Broadcast Center Function). Its main task is to modify or delete CBS messages, allocate serial numbers while indicating the geographical scope of each CBS message, initiate broadcast by sending fixed length CBS messages, determine the set of cells to which a CBS message should be broadcasted, determine the time at which a CBS message should commence or cease being broadcasted, and determine the period at which the broadcast of the CBS message should be repeated. Each CBC/CBCF may be connected to several AMFs or PWS-IWFs.

\noindent\textbf{PWS-IWF} (Public Warning System Interworking Function). The purpose of this logical function is to translate messages (\eg, Write-Replace-Warning-Indication and Stop-Warning-Indication) from N50 interface to SBc interface and vice versa. Finally, the PWS-IWF may interface to one or multiple AMFs and one or multiple CBCs.

\noindent\textbf{AMF} (Authentication and Mobility Function). In PWS the AMF provides reports and acknowledgements to the CBC/CBCF regarding the execution and forwarding of commands received from them, and routes the warning messages (\eg, Write-Replace-Warning Request) to the appropriate RAN nodes in the indicated Tracking Area. In addition, it reports the Broadcast Completed Area List, the Broadcast Cancelled Area List, the PWS Restart Indication and the PWS Failure Indication received from RAN nodes to all CBCs/CBCFs and PWS-IWFs that it interfaces with.

\noindent\textbf{NG-RAN} (Next Generation-Radio Access Network). It comprises gNodeBs and/or ng-eNodeBs which are the 5G related base stations. Upon reception of a command, it executes the associated procedure for the UEs in the target cells. For instance, a warning request will make the RAN deliver the proper paging messages to all UEs and then broadcast the SIBs as instructed. In the case of cancellation, the RAN ceases the transmission of warning messages. Finally, the RAN reports to the AMF regarding the execution of each command. 

\noindent\textbf{UE} (User Equipment). It is the mobile terminal of a subscribed user (with a dedicated USIM) that utilizes legitimate network services offered by a network provider.

The architecture mainly supports the CBCF as the CBC and PWS-IWF are considered optional entities.

\subsection{The Paging Procedure}

In cellular networks UEs enter into an RRC-Idle state to preserve battery when there is no active service or any ongoing data transmissions. When there is an upcoming service (\eg, incoming call) to be delivered to a specific UE, the AMF makes sure that the UE is in an RRC-Active/RRC-Connected state (if not already). By establishing an RRC Connection and the necessary radio bearers of data traffic a UE can have access to network services. In order to get this connection, UEs need to monitor for paging messages while in RRC-Idle or RRC-Inactive states at device-specific times and respond to the core network accordingly. This procedure is called \emph{Paging} and it is also used in PWS to warn users about emergencies. 

\begin{figure}[t]
    \centering
    \includegraphics[width=.85\columnwidth,keepaspectratio]{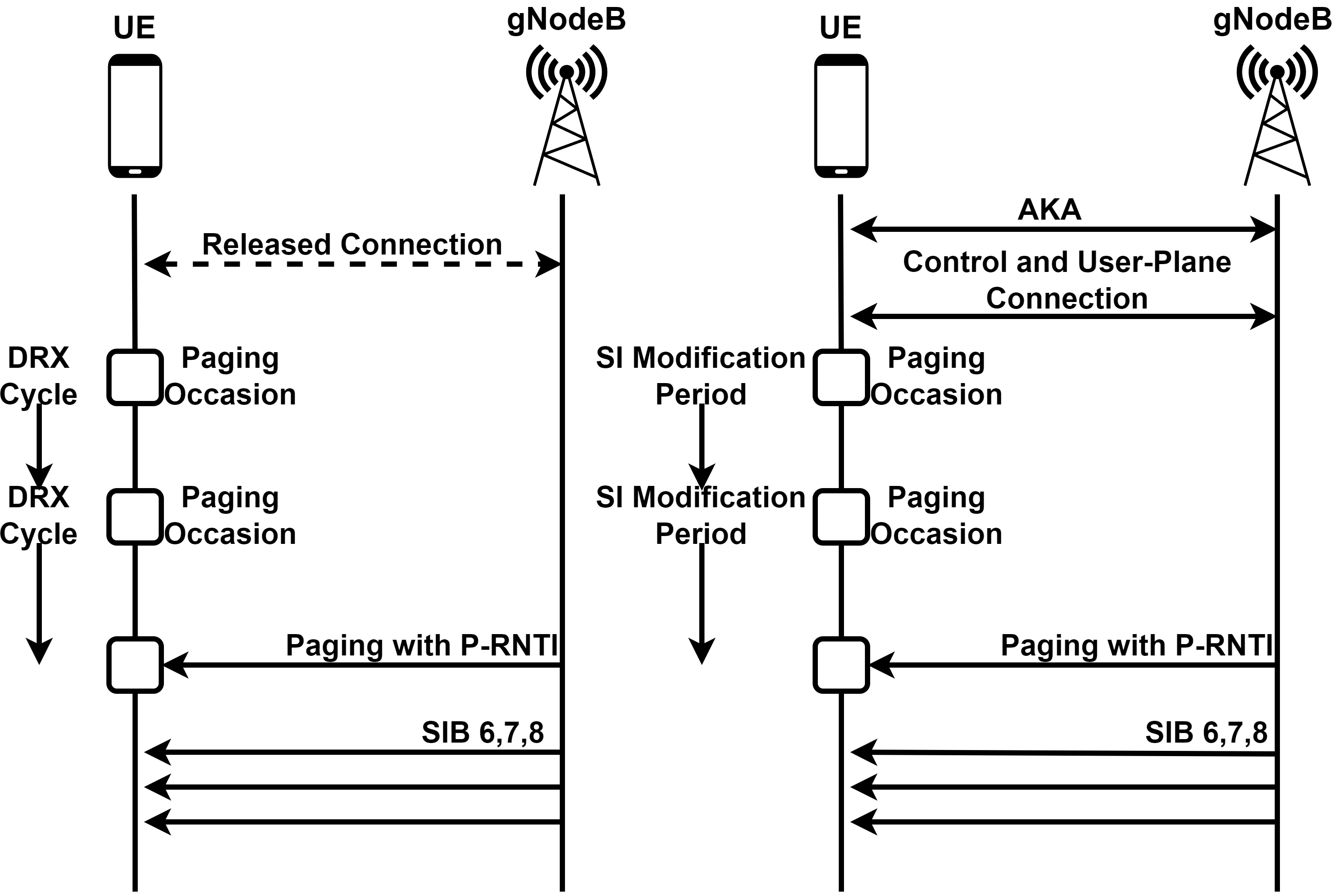}
    \caption{Warning procedure when the UE is RRC-Idle or Inactive (Left) and when in RRC-Connected state (Right)}
    \label{fig:warning_idle_inactive_connected}
\end{figure}

In PWS, ETWS/CMAS capable UEs in RRC-Idle or RRC-Inactive states monitor for indications about PWS notifications in their own paging occasion every Discontinuous Reception (DRX) cycle, whereas in RRC-Connected state the System Information (SI) Modification Period is used. Figure~\ref{fig:warning_idle_inactive_connected} shows how the paging procedure works. Specifically for 5G SA, the ETWS/CMAS paging procedure utilizes only the payload of the Physical Downlink Control Channel (PDCCH) with P-RNTI and a 'short message' in the Downlink Control Information Format 1\_0. Figure~\ref{fig:pwsgraph} in Appendix~\ref{Control-flow} presents the complete emergency flow on 5G SA.

\subsection{Broadcast and Warning Messages}

In PWS, the Core Network receives the warning messages and its configurations by the external entities. The Write-Replace-Warning Request contains all the necessary values to be considered by the AMF and sent to the RAN. The RAN translates the Write-Replace-Warning Request to the SIB messages that will be broadcasted. Finally, the RAN transmits paging messages to all associated cells with cause \emph{Emergency} and repeatedly broadcasts the SIB(s). UEs monitor warning indications in their own paging occasion for RRC-Idle and RRC-Inactive and in any paging occasion for RRC-Connected.
Warning types can be separated into two major groups: ETWS and CMAS, each having its own dedicated SIB. Figures~\ref{fig:sib6}--\ref{fig:sib8} (Appendix) show examples of SIB messages used during our experiments.
 
\emph{ETWS} is a PWS mechanism developed to meet the regulatory requirements for warning notifications related to earthquake and tsunami events. An ETWS warning notification can either be a primary notification (short notification) or a secondary notification (providing detailed information).
The ETWS Primary Notification, which is broadcasted by using SIB~6, carries small data to be sent quickly to the network and to indicate the imminent occurrence of earthquake and tsunami. The ETWS Secondary Notification, which is broadcasted by using SIB~7, carries a large amount of data in order to send text, audio (to instruct what to do), graphical data such as a map indicating the route from the present position to an evacuation site, etc. Furthermore, the ETWS Primary Notification has higher priority than the Secondary Notification, in case both notifications exist concurrently in a specific PLMN.

\emph{CMAS} is a PWS mechanism developed for the delivery of multiple concurrent warning notifications. These messages include CMAS Presidential Level Alerts, CMAS Child Abduction Emergency (\eg, AMBER), and Imminent Extreme or Severe Threats and Public Safety. SIB~8 is particularly assigned for CMAS messages. Some CMAS messages are always enabled (mandatory) in smartphones (shown in Figure~\ref{fig:alert-options} for the Huawei P40 5G test phone). 

Finally, Figure~\ref{fig:alerts-examples} (Appendix) shows an example of a CMAS message and an ETWS message in our experimentation.

\noindent \textbf{Warning Processing and Roaming.} PWS in roaming scenarios requires a separate treatment as a vital part of telecommunications. When a user enters a Visited Public Land Mobile Network (VPLMN), possibly in another country, the operator in the visited country is responsible for delivering warning messages in case of an emergency. Considering that both the Home Public Land Mobile Network (HPLMN) and VPLMN have set up their own PWS (otherwise the lack of a PWS can endanger the user), in roaming cases a PWS-capable UE needs to fulfill the requirements of the VPLMN's PWS service. This means that any incompatibilities between HPLMN and VPLMN should be eliminated.

%% file: sections/3.adversarial-setup.tex
\section{Adversarial Setup \& Weaknesses}

\subsection{Threat Model}

The attacker's ultimate goal is to wreak havoc among a population at maximum capacity by sending fake warnings or suppressing legitimate warnings to conceal an emergency. In our threat model, we consider an active adversary who has full protocol knowledge and the radio abilities to install and operate a base station with similar capabilities as a legitimate one. In particular, the fake station can mimic a legitimate base station and thus force a victim's device to connect to it by broadcasting spoofed Master Information Block (MIB) and System Information Block (SIB) messages in the victim's frequency. We make the standard assumption that the attacker is able to capture and craft MIB, SIB, paging and PWS CBS messages by eavesdropping the public channels which are broadcasted to the network users. In addition, we consider an attacker that can establish a MitM position between UEs and gNodeBs, which in turn may allow him/her to eavesdrop, drop, modify and forward messages while respecting the cryptographic assumptions. To carry out the attacks, he/she may utilize any available free or commercial equipment and setup multiple base stations. Finally, we assume that the adversary cannot have physical access to the USIM cards, mobile devices, RAN, or Core Network to obtain or alter sensitive information (\eg, cryptographic material) and we consider side-channel attacks as well as signal jamming as out of scope. 

\subsection{Frail Cellular Features and Flaws} \label{flaws}

\begin{figure}[tb]
    \centering
    \includegraphics[width=1.\columnwidth,keepaspectratio]{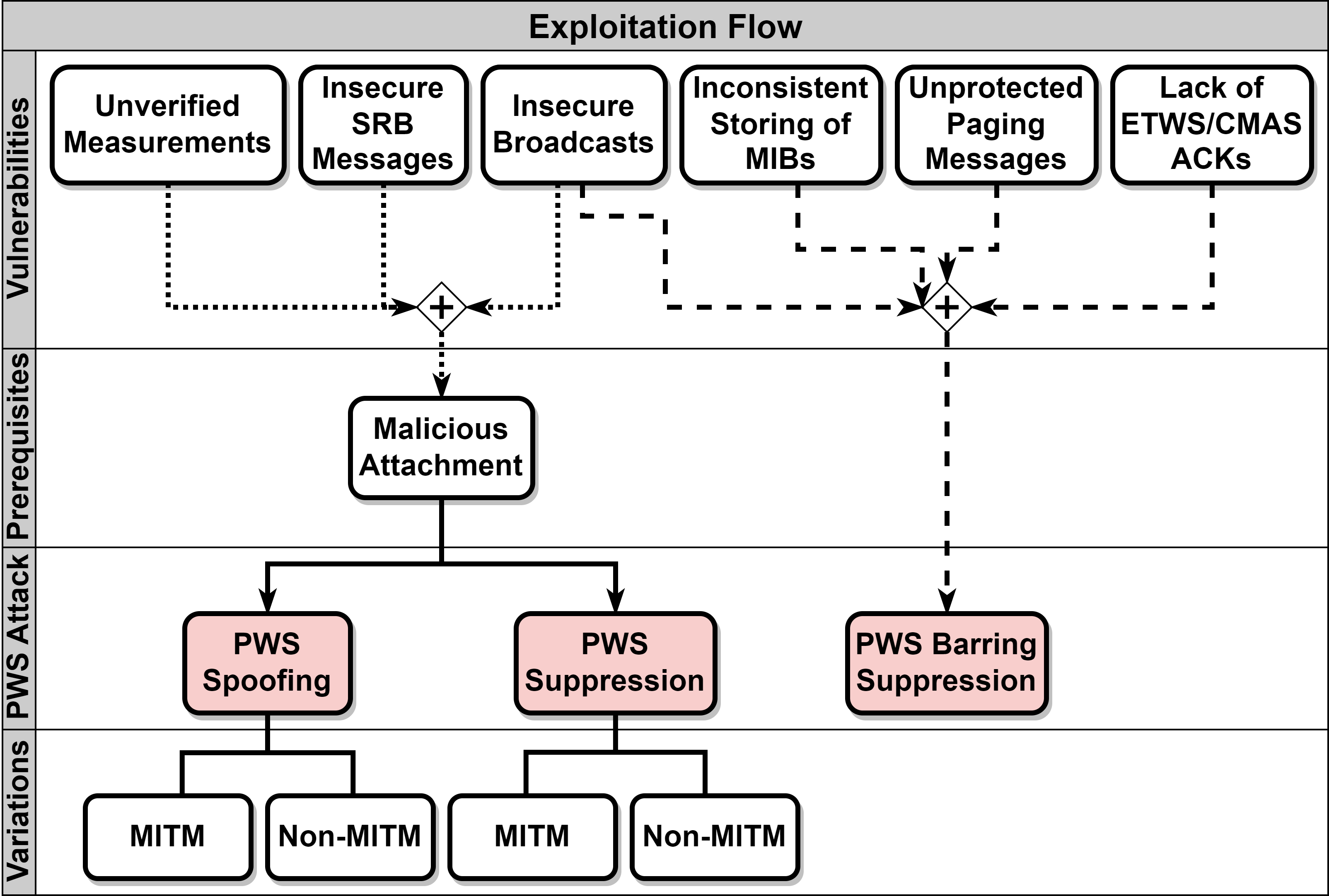}
    \caption{Exploitation flow, showing the connections between individual vulnerabilities and associated attacks, incl.~variations and prerequisites.}
    \label{fig:exp-flow}
\end{figure}

We identify and experimentally validate multiple security flaws that can be misused for PWS exploitation on the 5G SA domain. PWS exploitation consists of making a UE maliciously attach to the fake base station (phase 1, malicious attachment) and the actual PWS attacks being conducted (phase 2). Flaw 1 is used for both phases, flaws 2 and 3 for the malicious attachment only, and flaws 4-6 are associated with the PWS attacks. Figure~\ref{fig:exp-flow} shows which vulnerability contributes to each attack.

\noindent \textbf{(1) Insecure Broadcast Messages.} The MIB and SIB messages that are transmitted by legitimate base stations are used for UE attachment to the network and support of essential network operations (\eg, synchronization, handover, cell reselection procedures). However, these messages are not security-protected, being devoid of encryption, integrity-protection, and authentication. Thus, an attacker can capture the MIB and SIB messages and transmit them imitating real base stations (and cell(s)). The UE accepts the messages as there is no way to validate the source leading to malicious attachments. Specifically for SIB types 6, 7 and 8 that are related to 5G PWS, the UE receives the spoofed SIB-based warnings after a potential fake paging process and displays them to the user as normal, as long as the UE is attached to the attacker. We were able to verify that this weakness still exists on 5G in Sections~\ref{mal_attach} and~\ref{experiments}.

\noindent \textbf{(2) Unverified Measurements.} The UE is designed and instructed to monitor the network for the best possible signal quality and report its findings to the network. This signal quality concerns the efficiency of the mobility management since UE relocation from one cell to another becomes easier. However, any base station that broadcasts the MIB and SIB messages can make the UEs collect measurement data (\ie, RSRP, RSRQ, SINR), and a malicious base station can trick them. Moreover, a UE collects malicious measurements without any verification. As a consequence, the UE may use them to perform a cell reselection or handover~\cite{3gpp.123.502, 3gpp.138.304}. Typically, a Measurement Report is crafted and then sent to the RAN for evaluation. The RAN will accept the included measurements in the report without verification resulting in malicious handovers~\cite{Bitsikas21:handover-exploitation}, even though the Measurement Report is security-protected. Eventually, the UE relocates to the bogus base station which allows PWS manipulation. We illustrate this attack in Section~\ref{exploitation}. 

\noindent \textbf{(3) Insecure Signal Radio Bearer (SRB) Messages.} Apart from the potentially abused NAS messages, such as attach reject and service reject, Signal Radio Bearer 0 (SRB0) messages are not required to be sent securely according to the specifications~\cite{3gpp.138.331}. In addition, the RRC Release of the Signal Radio Bearer 1 (SRB1) can be transmitted and accepted without security protection. Thus an attacker can abuse these messages in order to exploit network users' RRC connections. The manipulation of these messages is apparent in past works on LTE~\cite{Kim19:touching-untouchables, Hussain18:LTE-Inspector} and 5G~\cite{Hussain19:5G-reasoner, Bitsikas21:handover-exploitation}. We confirm them and make them part of our PWS attacks. Such unprotected messages together can boost attacker's capabilities on traffic manipulation. In the context of PWS exploitation, the affected SRB messages can be used to expedite the malicious attachment to a false base station as the attacker can use them to manipulate UE's traffic,~\eg, leading to the establishment of a MitM relay to spoof or suppress alerts. 

\noindent \textbf{(4) Inconsistent storing of MIB messages.} MIB messages are used in order for the UEs to collect essential information about the network and decode the SIB 1 messages which are needed for the initial RAN connection. A UE searches for these messages and once it receives an MIB which is assigned to a specific cell of a base station, it follows a predefined set of instructions that determine if it must proceed with the connection or not. Furthermore, the UE stores the MIB before this decision until the smartphone reboots/shuts down or enters into an airplane mode wiping out its temporal memory. We discovered that an attacker can take advantage of this mechanism to make the UE store malicious MIB values ignoring the real MIBs while the UE remains functional, because the UE cannot accept new information about a certain cell without eliminating the old (malicious) first. We mainly use this inconsistency in our PWS barring attack where we make legitimate base stations look unavailable, since a UE is forced to store incorrect MIB information affecting the reception of warning messages. We explain this further in Section~\ref{attacks_non-mitm}.

\noindent \textbf{(5) Unprotected Paging Messages.} Paging messages lack cryptographic protection, and thus are susceptible to spoofing and forgery~\cite{Hussain19:privacy-attacks-paging, Hussain18:LTE-Inspector}. Even though security enhancements have been considered and implemented~\cite{3gpp.138.331, EricssonReport} on 5G SA; temporary identifier usage (5G-TMSI or I-RNTI) instead of permanent, removal of long-term permanent paging identifier, robust randomization and frequently refreshing the temporary identifiers, the lack of integrity-protection and authentication render the aforementioned defenses inadequate for PWS cases. Specifically, we reveal through 5G experimentation that 5G suffers from the same security flaw as LTE~\cite{Hussain18:LTE-Inspector}. To be more exact, the attacker can send fabricated PWS-based paging messages when necessary along with the malicious SIB 6, 7 or 8 broadcast messages. Furthermore, paging messages are designed to include the 16-bit fixed P-RNTI value $65534$ ($0xFFFE$)~\cite{3gpp.138.321, 3gpp.138.331} for all UEs in the targeted Tracking Area. We verified that this feature is problematic as the attacker circumvents all the aforementioned countermeasures and does not require any type of sniffing to collect temporary identifiers for each UE in the area. As a consequence, the attack becomes less convoluted to execute.

\noindent \textbf{(6) No  Acknowledgements in ETWS/CMAS Delivery.} The paging procedure and SIB transmission mechanism lack acknowledgements from the corresponding UEs. The UE only receives the alerts and afterwards displays the warning message to the user. However, the Core Network does not know if a particular or any UE in a Tracking Area has received the warning message. The UE receives the paging message in a paging occasion and the associated SIB messages, but does not respond back to the gNodeB (see Fig.~\ref{fig:warning_idle_inactive_connected}). We verified through experimentation that this may instill implications in the PWS mechanism as an attacker can leverage this weakness to make spoofing and suppression attacks less discernible to the operator. Finally, since the core network may collect traces of successful or failed warning distributions for evaluation and error correction (last step in App.~\ref{Control-flow}), these procedures may not be accurate.

%% file: sections/4.exploitation.tex
\section{Exploiting the PWS}\label{exploitation}

We now break down each attack variation and detail each execution.

\subsection{Malicious Attachment} \label{mal_attach}

The first phase of the PWS spoofing and suppression attacks comprises the malicious attachment of the victim UE to the attack equipment: The attacker attracts UEs to connect to the false base station by satisfying the signal threshold requirements, while forcefully breaking any connection with the legitimate network. To accomplish this, the attacker sets up a false base station (Sec.~\ref{setup}). Chances of success are better if the replayed \texttt{cell\_reselection\_priority} of SIB type 2 has the maximum value (\ie, $7$).

To be specific, the UE will get maliciously attached to the fake station depending on the RRC states it is in when the attack starts:
\begin{itemize}
\item If the UE is in RRC-Idle state,  \emph{cell selection} and \emph{reselection} happen. In the case of an RRC-Inactive state, where the UE has a suspended connection, it might be necessary to transition to the RRC-Idle state first with a connection release and then perform the procedure above.  
\item If the UE is in RRC-Connected state, reports false malicious measurements in the Measurement Report and passes the signal strength threshold, the \emph{handover} procedure (Xn or N2) will happen. The handover procedure is executed without any verification by the RAN. Even though the handover may eventually fail on a network, once the UE receives the RRC Connection Reconfiguration, it attaches to the malicious cell.
\end{itemize}

Figures~\ref{fig:attackgraph1} and~\ref{fig:attackgraph2} demonstrate the interrupted communication which corresponds to the detachment (step 1) and then the connection to the rogue base station. In step 2, the attacker needs to respond to the victim with the proper SRB 0 and 1 messages. The process typically begins with an RRC Reestablishment Request (with cause \texttt{handover\_Failure}) or RRC Setup Request by the UE to recover the previous connection or start anew, respectively. The attacker should respond with an RRC Reject in case of reestablishment as he/she cannot offer legitimate services and does not possess the cryptographic keys. This will turn the disrupted connection into a fresh one, compelling the UE to setup a new RRC connection. In case the UE sends the RRC Setup Request at the beginning instead, the attacker should permit the RRC connection if possible. It is also probable that the UE sends a Service Request no matter the case. The attacker needs to send back a Service Reject and then an RRC Release for the same reasons as in reestablishment situations. Eventually, the UE initiates an RRC connection again and then sends the NAS Attach Request to the attacker. The attacker can either forward the request to the legitimate network along with the subsequent traffic and setup a MitM relay or reject it continuously until the UE fully disconnects.  Appendix~\ref{setup} provides more information about the false base station setup.

\subsection{Attacks based on MitM} \label{attacks_mitm}

\begin{figure}[tb]
    \centering
    \includegraphics[width=1.\columnwidth,keepaspectratio]{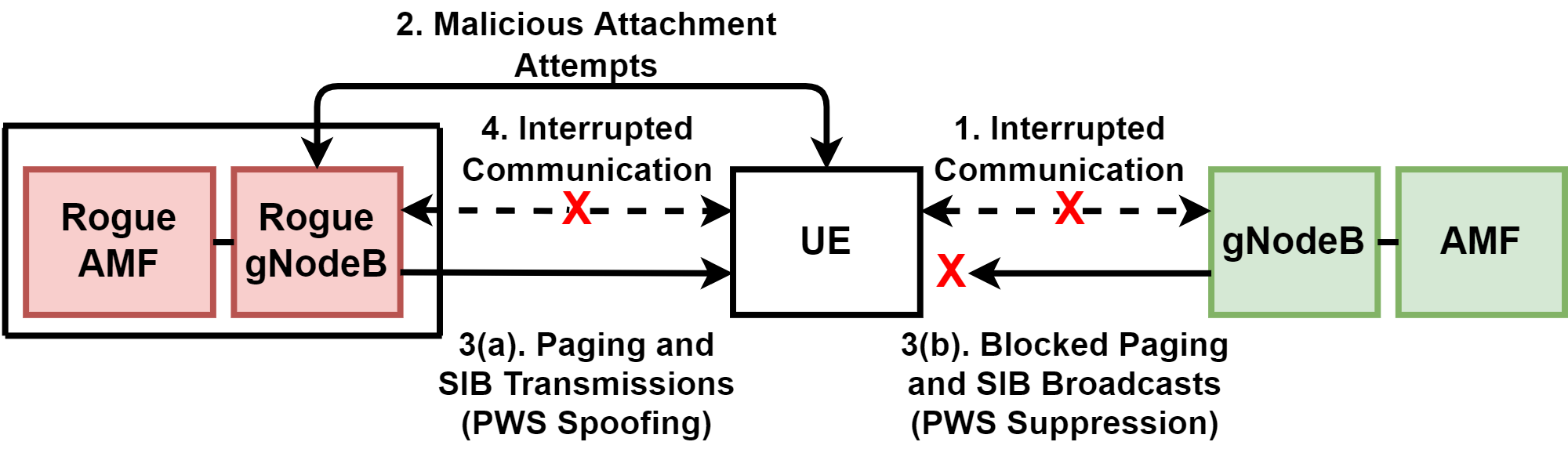}
    \caption{Spoofing and Suppression on a non-MitM Setup}
    \label{fig:attackgraph2}
\end{figure}

PWS suppression and spoofing attacks are possible in a MitM setup, see Figure~\ref{fig:attackgraph2}. The MitM setup can be established through a cell (re)selection or a handover procedure similar to~\cite{Bitsikas21:handover-exploitation, rupprecht19:layer-two, rupprecht20:imp4gt, Hussain19:5G-reasoner}.

\begin{figure}[tb]
    \centering
    \includegraphics[width=1.\columnwidth,keepaspectratio]{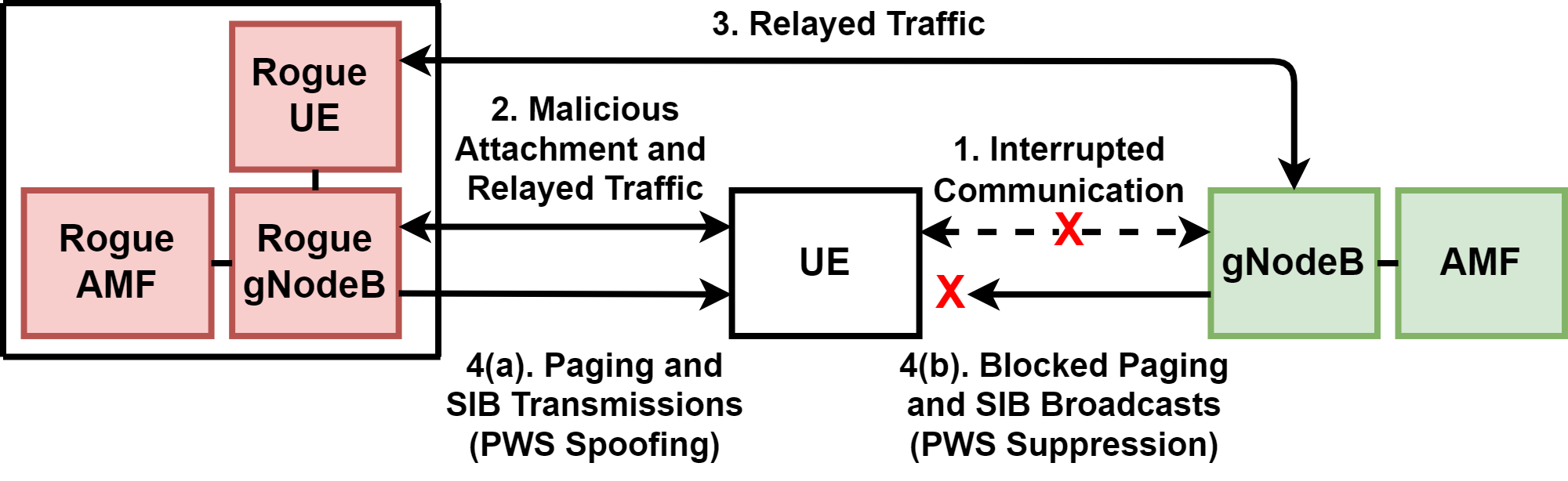}
    \caption{Spoofing and Suppression Attacks on a MitM Setup}
    \label{fig:attackgraph1}
\end{figure}

\textbf{PWS Spoofing Attack}: Based on the attachment of the UE to the false cell and given that the attacker has replayed the NAS Attach Request to the real network with all the subsequent up- and downlink traffic (step 3 in Fig.~\ref{fig:attackgraph1}), the attacker is in a MitM position, allowing them to exploit the PWS. The actual exploitation unfolds when the attacker forges and transmits fake warning (CMAS \& ETWS) messages for all paging occasions. Since the UE believes it communicates with a legitimate base station, it accepts all warning messages without verification. The UE is locked to this bogus cell, %meaning that it 
accepting warning messages only from it as long as it stays connected, even though the real cell may transmit other messages. Figure~\ref{fig:attackgraph1} shows that the attacker sends PWS-based paging messages to keep the UE in RRC-Connected state along with the SIB broadcasts with maximum periodicity (step 4a). As long as the UE remains locked without disruption, it receives the malicious alerts.

Nevertheless, the \emph{spoofing duration}, $D_{spoof}(MitM)$, which we define as the time of the UE between starting the RRC Re\-establishment or RRC Setup of the malicious attachment after a potential RACH process until the total disconnection from the attacker, is not static, since the malicious connection may fail and/or the UE may break away entering into a DoS state. Fluctuations in the duration may also depend on the smartphone device (due to different baseband implementations) and potentially disrupted services (Call, SMS, Internet Data, etc.) before the malicious attachment. Once the UE disconnects, the attacker can no longer spoof warnings, especially if the UE evades the attacker's range. Thus, contrary to what is reported in~\cite{Lee19:spoofing-alerts-lte, Jihoon21:Securing-Wireless-Alerts}, PWS spoofing is also possible through handover exploitation when the attacker imitates a legitimate base station and when a MitM is established.

\textbf{PWS Warning Suppression}: Suppressing genuine warning messages is possible through detachment from legitimate base stations and then malevolently connecting to a false base station. In this case, the UE is locked to the attacker's station overlooking legitimate services. In Figure~\ref{fig:attackgraph1}, the UE is not receiving the paging and warning-based SIB messages when attached to the false cell (step 4b). The network believes that warnings have been delivered successfully, however the lack of acknowledgements and untriggered PWS Failure Indication makes the attack less detectable. The attacker can continue relaying traffic as normal and even spoof at the same time with the legitimate network. The suppression continues until the UE disconnects from the attacker and connects to the real network appropriately. The disconnection may occur due to connection failures or explicitly by the attacker (\eg, through NAS Detach Request). Our experimentation showed that the UE cannot recover unless airplane mode or rebooting is used when the UE enters into a DoS state. Therefore, legitimate warning notifications cannot be received and displayed to the user at that time.

Thereupon, we can estimate the aggregated \emph{Suppression Duration} for a specific UE-victim as: 
\begin{equation} \label{algo:suppress-mitm}
\begin{split}
    D_{supp}(MitM) & \approx D_{spoof}(MitM) + t_{rec,supi} + t_{rach,ran}
\end{split}
\end{equation}
\noindent where the $D_{spoof}(MitM)$ is the spoofing time in a MitM setup till the UE disconnects, the $t_{rec,supi}$ is the recovery time of the UE device with a specific SUPI, and $t_{rach,ran}$ is the time it takes for the UE to find the legitimate RAN and complete a RACH procedure while beginning the RRC message exchange. 

\subsection{Attacks Without MitM} \label{attacks_non-mitm}

The attacker does not need to perform any message relay, but can respond to the UE until the connection breaks~\cite{Lee19:spoofing-alerts-lte, Shaik18:impact-SON}. Specifically, after multiple attachment attempts fail, the UE abandons the malicious attachment and becomes deregistered.

\textbf{PWS Spoofing Attack}: Similar to MitM cases, the spoofing takes place once the UE connects to the bogus cell. This can happen either through a handover procedure or a cell (re)selection that will make the UE send the RRC and NAS messages (Sec.~\ref{mal_attach}). When the UE transmits the NAS Attach Request, the attacker repeatedly responds with a NAS Attach Reject (step 2 in  Fig.~\ref{fig:attackgraph2}). The UE tries several times to establish a connection without any fruitful outcome. On the attacker's side, the spoofing takes place starting from the RRC Reestablishment or RRC Setup as in the previous scenario. Moreover, the spoofing continues throughout the entire attachment process (step 2) with maximum transmission since once again the UE accepts all CMAS and ETWS warning messages sent by the attacker without validation. Eventually, once the UE stops pursuing the attachment, it disconnects and the attacker ceases the attack (step 4). The UE enters into a DoS state until it recovers.

The \emph{spoofing duration}, $D_{spoof}(Attach)$, starts from the RRC Reest\-ablishment or RRC Setup as in the MitM setup, but ends with the last Attach Reject of the attacker which forces the UE to disconnect. This means that the duration is shorter compared to the $D_{spoof}(MitM)$, because it depends on UE's tolerance on failed attachments (typically $5$ times). Even though, the spoofing duration is reduced considerably, this type of attack is less complicated since it does not necessitate the traffic to be relayed to the real network. Therefore, the trade-off here is less complexity for less attacking impact.

\textbf{PWS Warning Suppression}: Suppression in this scenario happens throughout the malicious attachment as the UE does not have a connection with the legitimate network in order to receive paging and warning notifications (step 3b in Figure~\ref{fig:attackgraph2}). Similar to the MitM cases, the lack of acknowledgements and security-related indications in the PWS can make the attack less detectable. Once the UE receives the last NAS Attach Reject, it totally disconnects, and will be unable to receive warning notifications even if the malicious attachment ceases (step 4). Recovering will require the user to reboot the device or utilize the airplane mode. Hence once again, the \emph{suppression duration} can be approximated as follows:
\begin{equation} \label{algo:suppress-nonmitm}
\begin{split}
    D_{supp}(Attach) & \approx D_{spoof}(Attach) + t_{rec,supi} 
    + t_{rach,ran}
\end{split}
\end{equation}
\noindent where the $D_{spoof}(Attach)$ is the spoofing time in a non-MitM setup as a simple malicious attachment until the UE disconnects.

\begin{figure}[t]
    \centering
    \includegraphics[width=0.73\columnwidth,keepaspectratio]{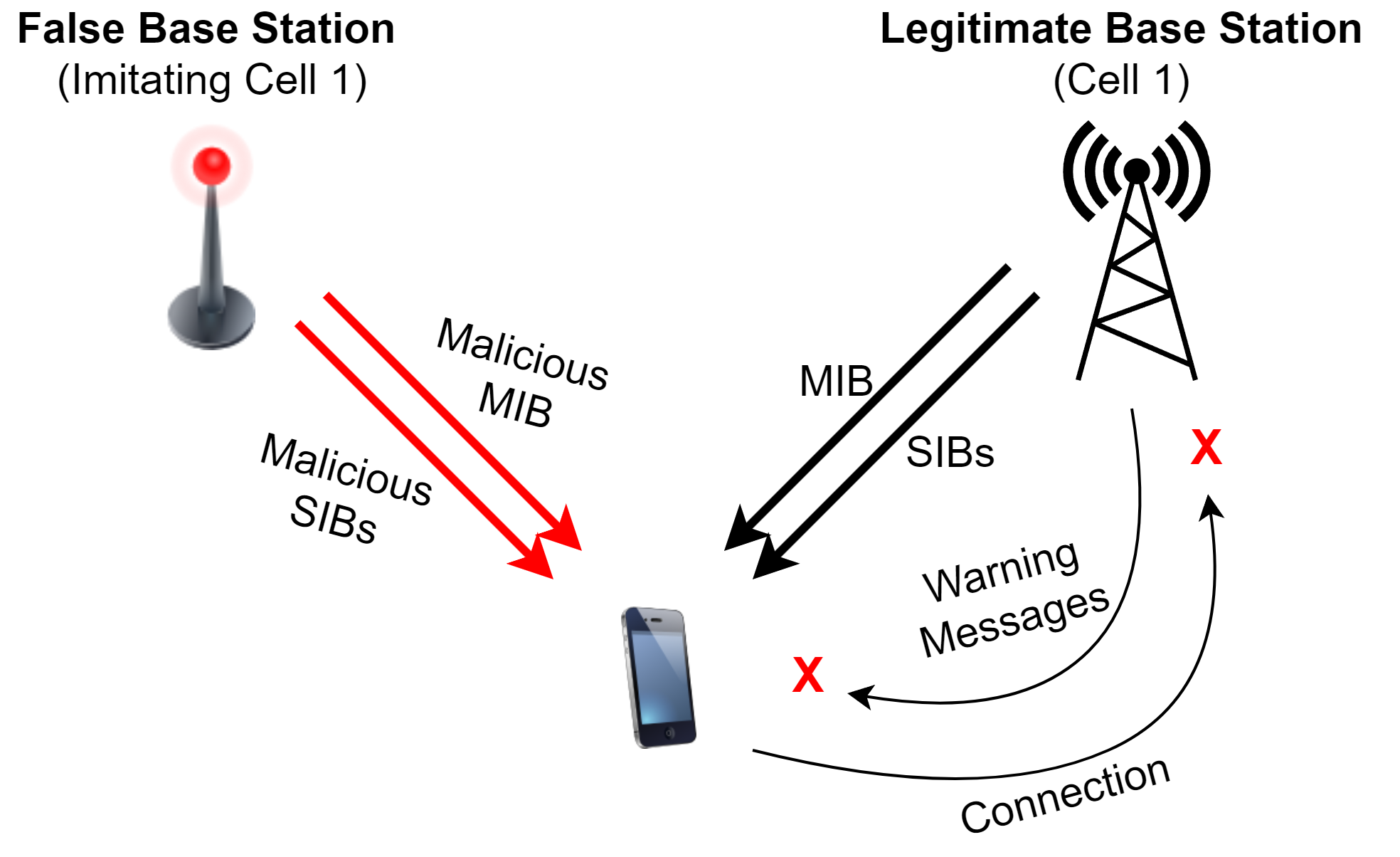}
    \caption{Our Barring Attack for Warning Suppression}
    \label{fig:barrattack}
\end{figure}

\textbf{PWS Barring Attack}: This type of attack is an independent case that does not demand a malicious attachment and a MitM setup. The goal is to disallow any connection to a legitimate base station, thus suppressing the warning messages that are destined for a specific cell/Tracking Area. The barring attack takes advantage of 5G access control, MIB/SIB storage mechanism and lack of MIB/SIB security, and manipulates the MIB and SIB type 1 messages. Once the adversary commences the transmissions, the UEs receive the malicious broadcast messages and decide not to connect to the legitimate base station, as shown in Figure~\ref{fig:barrattack}. 

Like in the previous attacks, the attacker will need to configure the base station as the legitimate one, therefore capturing the MIB and SIB broadcasts is necessary. Nevertheless, the key element of this attack is the modification of three parameters instead of just replaying the captured messages: (1) Set \texttt{cell\_barred} of MIB to 'barred', (2)  \texttt{intra\_freq\_reselection} of MIB to 'notAllowed', and (3) \texttt{cell\_reserved\_for\_operator\_use} of SIB 1 to 'reserved'. Typically, these fields are used for maintenance, private access, and other operational purposes by the operator. We choose to modify SIB 1 as well in order to bolster the efficiency of our attack, even though the MIB is sufficient on 5G SA. We found that other fields, such as the \texttt{cell\_reselection\_priority} in SIB 2, are not necessary to abuse, as the UE processes the MIB and SIB 1 first.

In 5G, the \texttt{cell\_barred} parameter allows early detection of the cell's status without requiring the UE to receive and decode the SIB 1. If the MIB indicates that a cell is barred then the UE will also check the \texttt{intra\_freq\_reselection} parameter; a flag of ‘notAllowed’ indicates that the UE is not permitted to reselect another cell on the same frequency. The UE typically has to wait 300 seconds before re-checking this MIB to determine whether or not this cell remains 'barred'. Consequently, this allows early suppression of the warning messages. On the contrary, in LTE, both above fields are located in SIB 1 instead, which follows the MIB. Finally, \texttt{cell\_reserved\_for\_operator\_use} could be broadcasted with a value of ‘reserved’. Then, a UE with an Access Identity of $11$ (PLMN Use) or an Access Identity of $15$ (PLMN Staff) is allowed to use the cell for selection and reselection only, while a UE with Access Identity $0$ (no configuration), $1$ (Multimedia Priority Service), $2$ (Mission Critical Service), $12$ (Security Services), $13$ (Public Utilities) or $14$ (Emergency Services) treats the cell as ‘barred’, prohibiting selection and reselection.

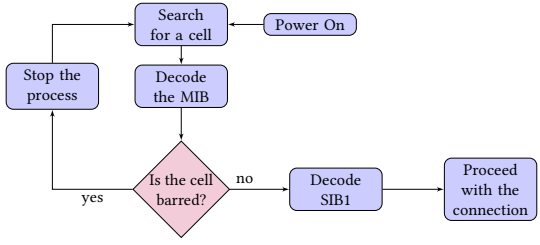
\begin{figure}[t] 
\centering
% Define block styles
\tikzstyle{decision} = [diamond, draw, fill=purple!20, 
    text width=4em, text badly centered, node distance=3cm, inner sep=0pt]
\tikzstyle{block} = [rectangle, draw, fill=blue!20, 
    text width=5em, text centered, rounded corners, minimum height=1em]
\tikzstyle{line} = [draw, -latex']
\resizebox{0.4\textwidth}{!}{%
\begin{tikzpicture}[node distance = 1.2cm, auto]
    % Place nodes
    \node [block] (init) {Power On};
    \node [block, left of=init, node distance=2.5cm] (search) {Search for a cell};
    \node [block, below of=search] (mib) {Decode the MIB};
    \node [block, left of=mib, node distance=2.5cm] (stop) {Stop the process};
    \node [decision, below of=mib, node distance=2cm] (decide) {Is the cell barred?};
    \node [block, right of=decide, node distance=3cm] (sib) {Decode SIB1};
    \node [block, right of=sib, node distance=3cm] (proceed) {Proceed with the connection};
    % Draw edges
    \path [line] (init) -- (search);
    \path [line] (search) -- (mib);
    \path [line] (mib) -- (decide);
    \path [line] (decide) -| node [near start]{yes}(stop);
    \path [line] (stop) |- (search);
    \path [line] (decide) -- node [near start]{no}(sib);
    \path [line] (sib) -- (proceed);
\end{tikzpicture}
}
\caption{Access Control Process and Cell Connection}
\label{fig:access-control} 
\end{figure}

Furthermore, as indicated by the inconsistent storing of MIB messages (flaw 4 in Sec.~\ref{flaws}), broadcast reception and storing processes can be erroneous. Typically, the UE stores the first MIB instance as it follows a predetermined set of instructions. Consequently, it may ignore other instances and reject legitimate MIBs, thus never decoding the legitimate SIB 1 in order to connect to the corresponding real cell. This set of instructions is presented in Figure~\ref{fig:access-control}, clarifying that in case of a malicious MIB, the UE will never proceed to SIB 1 decoding altogether. If the UE has no saved information of the targeted cell and no connection has been established (at least a completed RACH), it is highly possible that it will accept and process the malicious MIB and SIB transmissions. Additionally, even if the legitimate base station transmits its own versions of broadcast messages simultaneously, the UE will overlook them and comply with the bogus ones, if the false base station's signal strength is dominant. The attack cannot succeed, though, if the UE has already attached to the cell, since the attacker does not have a way to delete the stored information within the UE directly, possibly only through other attacks (\eg, DoS with detachments) that can force reset prior to launching the barring attack.

Given the cell gains of the legitimate station, $g_i$, and of the malicious station, $g'_i$, where $g_i,g'_i \in [-120dB, 0dB]$, their difference $\delta_i$ can be calculated: $\delta_i = |g_i - g'_i|$. In our experimental setup we discovered that the attack succeeds ($\alpha = 1$) when $\delta_i\geq10 dB$ and fails for any other condition in our setup:
\begin{equation} \label{algo:decision}
  \alpha =\begin{cases}
    1, & \text{if $\delta_i\geq10 dB$}.\\
    0, & \text{otherwise}.
  \end{cases}
\end{equation}
Signal strength is enough to ensure that the message will be received by the victim without dealing with the order of message reception or broadcast periodicity rendering the attack even more trivial to perform. In real-life scenarios, the signal strength needs to be adapted accordingly. 

This kind of suppression disrupts cell selection, reselection, and handover procedures, as the UEs will consider the affected cell as unavailable/blacklisted, leading to DoS and handover/reselection failures. Most importantly, the UE is unable to receive warning messages since attachment to the network is not feasible. It will be able to have normal services again when the attacker ceases the malicious transmissions or the UE escapes the attacker's range to connect to another available cell. This means that the barring attack starts from the decision that a cell is barred during the access control procedure until the attack stops or the UE evades the attacker's coverage. In other words, the \emph{Suppression Duration} $D_{supp}(Barr)$ is:
\begin{equation} \label{algo:suppress-barr}
\begin{split}
    D_{supp}(Barr) \approx t_{barr} + t_{rec,supi} + t_{rach,ran}
\end{split}
\end{equation}
\noindent where $t_{barr}$ is the time from the barring decision until the start of the disconnection.

%% file: sections/5.experimentation.tex
\section{Experimentation} \label{experiments}

We conducted a thorough practical evaluation of the presented attack on a set of smartphones. 

\subsection{Experimental Setup}

\begin{figure}[t]
    \centering
    \includegraphics[width=.75\columnwidth,keepaspectratio]{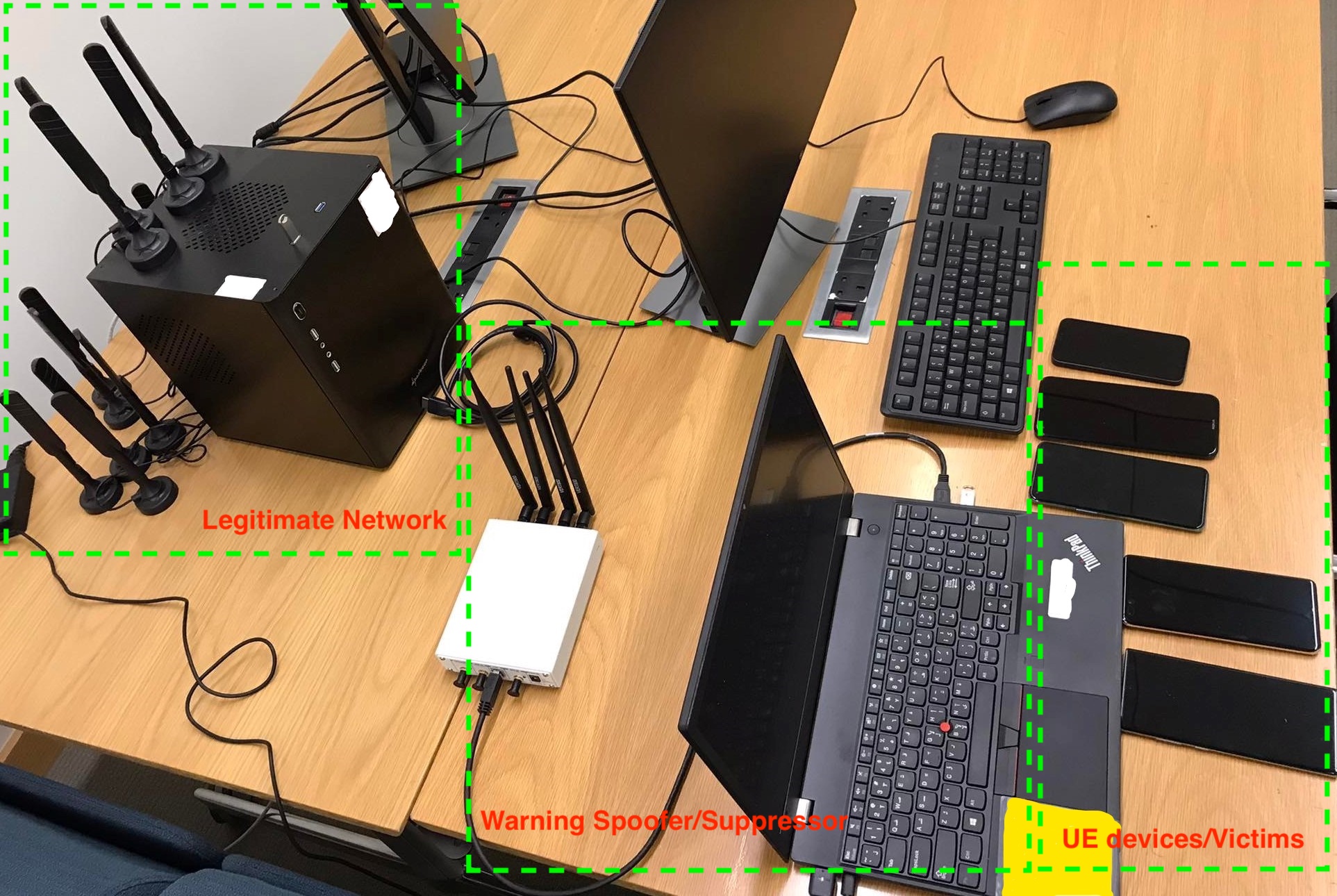}
    \caption{Our Experimental Setup}
    \label{fig:setup}
\end{figure}

Our setup comprises an Amarisoft Callbox Classic (equipped with SDRs)~\cite{amarisoft} with the 5G Core Network and the gNodeB representing the legitimate network (Figure~\ref{fig:setup}). Additionally, we have a Lenovo Thinkpad T580 laptop with Ubuntu 20.04 and an Ettus B210 USRP~\cite{ettus} for the malicious base station (with an approximate cost of 2k€). In our setup, we utilized the Amarisoft software for all 5G cases with a Core Network and a single gNodeB. In addition, we used numerous smartphone devices that were 5G and PWS-capable with an Anritsu SIM card. Table~\ref{tab:specs} shows the specific devices that we employed for 5G SA and NSA testing. More details about the exact cellular network configurations are presented in Appendix~\ref{network_conf}. We used the \texttt{cell\_gain} command with a maximum value of zero to trigger malicious attachments and handovers between cells.

For the MitM setup (Section~\ref{attacks_mitm}), our goal was to keep the victim attached to the rogue base station by responding to it normally without the need for further exploitation (\eg, RRC and NAS message modifications). Unfortunately, due to the black-box and commercial nature of Amarisoft software we could not establish a full-scale MitM as it would require minor architectural modifications that are usual for an attacker's setup, as in~\cite{rupprecht19:layer-two, rupprecht20:imp4gt}. This was not an issue for our attacks though, as we sufficiently used another identical AMF (reachable but not controlled by the attacker) in order to respond to the victim-UE accordingly.

\begin{table}[bt]
\centering
\caption{Device Specifications and Results. PWS Spoofing (Spoof) and Suppression (Supp) succeeded on all devices.}
\Large
\resizebox{\columnwidth}{!}{%
\begin{tabular}{@{}l|llll|c}
\textbf{Device} & Chipset  & OS & Model & Release & PWS  \\
& & & & & Spoof / Supp\\ \hline \hline
\\[\dimexpr-\normalbaselineskip+2pt]
\textbf{Huawei} & Huawei & Android & ELS-NX9 & 2020 & \cmark / \cmark \\ 
\textbf{P40 Pro 5G} & Kirin 990 5G & 10 & &\\ 
\hline
\\[\dimexpr-\normalbaselineskip+2pt]
\textbf{Nokia} & Snapdragon & Android & TA-1243 & 2020 & \cmark / \cmark \\ 
\textbf{8.3 5G} & 765G 5G & 10 & &\\ \hline
\\[\dimexpr-\normalbaselineskip+2pt]
\textbf{One Plus} & MediaTek Dimen-& Android & DN2101 & 2021 & \cmark / \cmark \\
\textbf{Nord 2 5G} & sity 1200 5G & 11 & &\\ \hline
\\[\dimexpr-\normalbaselineskip+2pt]
\textbf{Apple} & Qualcomm & iOS & MGDX3AA & 2020 & \cmark / \cmark \\ 
\textbf{iPhone 12 mini} & X55 modem & 14.1 & & \\ \hline
\\[\dimexpr-\normalbaselineskip+2pt]
\textbf{Samsung} & Snapdragon & Android & SM-N976Q & 2018 & \cmark / \cmark \\
\textbf{Note 10 5G} &  845 & 10 & & \\
\end{tabular}%
}
\label{tab:specs}
\end{table}

Figures~\ref{fig:sib6}-\ref{fig:sib8} show examples of the SIB warning structures that we used. The \texttt{message\_} \texttt{Identifier} field in SIB 6, 7 and 8 respectively shows the 16-bit value in hexadecimal that has to be included in each message. For ETWS we used the ID $1102$. For CMAS messages we used the ID range from $1112$ to $111B$ (HEX), where $1112$ is dedicated to  Presidential alerts, $1113$ to $111A$ to Extreme and Severe alerts, and $111B$ to Amber Alerts. In our experiments, the serial number of warning messages was between $0x3000$ and $0x5000$. The associated paging messages that were generated are presented by Figure~\ref{fig:paging}. Appendix~\ref{warning_conf} provides more details about our warning structure. Finally, Figures~\ref{fig:warningflow} and \ref{fig:sibflow} show the warning flow between the legitimate network entities for several attempts and a part of its physical layer transmissions respectively in our setup.

\textbf{Ethical Considerations.} The experiments were carried out in a confined lab testing environment without affecting legitimate services and real operators. To cancel any interference, we ensured that the experimentation range remained within 10 meters and we configured the setup with our own network and warning values, dissimilar to legitimate local networks and users. Other smartphone devices (w/o SIM) that were attached on real commercial operators were not affected during our experiments.

\subsection{Experimental Results} 

PWS attacks are applicable to all users, regardless of owning a SIM card, since real-world access to the emergency services is typically unrestricted. In Table~\ref{tab:attack-results} we present the attack variations and an empirical rating in terms of complexity and impact. For \emph{impact}, we primarily consider the maximum attacking duration of each variation, whereas for \emph{complexity}, we take into account the setup requirements, the traffic (re)direction of the attack, the necessary signal strength, and the preparation steps before the attack (\eg, broadcast messages modifications, RRC and NAS capabilities, \etc). \emph{High} complexity was assigned if thorough preparation, malicious attachment, traffic rerouting, and higher signal strength were required for a successful attack, whereas \emph{medium} and \emph{low} complexity describe those cases where attachment without traffic handling or no attachment at all, respectively, were sufficient. \emph{Impact} depends on the attack duration: the smaller the duration is, the lower the impact is scored. To assign an impact score we only compare variations of the same type of attack (either spoofing or suppression).

\begin{table}[tb!]
\centering
\caption{\label{tab:attack-results} Results for each attack. We evaluate each attack on a [Low, Medium, High]-scale according to our experiments and real-life adaptations including their approximate attacking durations in seconds. For the PWS barring attack, there is no specific lower and upper bound.}
\resizebox{\columnwidth}{!}{%
\begin{tabular}{@{}l|l|l|l@{}}
\toprule
\textbf{PWS Attack} & Complexity & Impact & Attack Duration (s)\\ \hline \hline
\\[\dimexpr-\normalbaselineskip+2pt]
\textbf{Spoofing (MitM)} & \shortstack{High} & \shortstack{High} & $D_{spoof}(MitM) \geq 55$\\
%\hline
\\[\dimexpr-\normalbaselineskip+2pt]
\textbf{Spoofing (non-MitM)} & \shortstack{Medium} & \shortstack{Low} & $D_{spoof}(Attach) \leq 43$\\ 
\hline
\\[\dimexpr-\normalbaselineskip+2pt]
\textbf{Suppression by DoS (MitM)} & \shortstack{High} & \shortstack{Medium} & $D_{supp}(MitM) \geq 58$\\ 
%\hline
\\[\dimexpr-\normalbaselineskip+2pt]
\textbf{Suppression by DoS} & \shortstack{Medium} & \shortstack{Low} & $D_{supp}(Attach) \leq 46$ \\
$\quad $\textbf{(non-MitM)} & &  & \\
%\hline
\\[\dimexpr-\normalbaselineskip+2pt]
\textbf{Suppression by barring} & \shortstack{Low} & \shortstack{High} & $D_{supp}(Barr) \in \mathbb{Q}^+$ \\
\bottomrule
\end{tabular}
}
\end{table}

Even though the impact of MitM-based attacks is higher due to a potentially long spoofing duration, the complexity also increases as the attacker needs a robust system able to establish and handle the UE connection with a legitimate cell, an arduous task in real-life scenarios. In our experiments, we were able to maintain at least a $D_{spoof}(MitM) \geq 55$ sec, which is longer than the duration in non-MitM cases ($\approx 40-43$ sec) allowing a $D_{supp}(MitM) > D_{supp}(Attach)$ as well. The approximate duration in non-MitM cases could also depend on the \texttt{emm\_cause} of rejections (\eg, \texttt{UE identity cannot be derived by the network} or \texttt{Implicitly detached}) and the manufacturer. Oppositely, attacks that do not rely on MitM setups are less complex since they only respond to UEs without consuming resources to manage and redirect traffic. The impact in these cases is, however, significantly reduced since the UE ceases the malicious attachment after a few attachment attempts. Finally, the PWS barring attack achieves high impact with low complexity due to its trivial setup, lack of traffic handling and large attacking duration. In our setup, we noticed that for 100\% success rate the barring attack requires less signal amplification, $\delta_i\geq10 dB$, than malicious MitM and non-MitM attachments, $\delta_i\geq30 dB$ (PWS barring can achieve approximately 90\% success rate for $5 dB$.).

Table~\ref{tab:spoofing-conf} presents our tested PWS configurations that could be used to magnify spoofing. Although the category \emph{sufficient impact} can achieve successful spoofing,  \emph{maximum impact} is more reliable in preserving a high-rate dissemination of alerts and in reaching more UEs. Our PWS attacks had the same effect on all smartphone devices from Table~\ref{tab:specs} during our experimentation. We also emphasize that advanced Amarisoft boxes can handle up to $1000$ devices, meaning that constrained and dense areas can be impacted severely when the attacker employs sophisticated and multiple FBSs in precise locations. In this sense, the attack can be expected to scale, despite remaining a local attack. Appendix~\ref{further-discussion} provides further considerations of the impact.

\noindent\textbf{Impact on IMS Emergency Calls.} We noticed that suppression can cause severe implications against the IMS Emergency Call Support, disallowing the user from using VoNR emergency calls (\eg, $911$ using SIP) on a 5G-capable PLMN when attached to the false cell. Since the UE is maliciously attached or suppressed through barring, IMS messages (\ie, Register, Subscribe, Notify and PRACK)~\cite{3gpp.23.128} along with RRC Reconfiguration and Session Modification messages are unattainable, thus call preparation will not occur. This is possible even without the use of \texttt{ims-EmergencySupport5GC} as false in SIB type 1 by the attacker. In fact for barring attacks, the attacker can accomplish this without any further change in the configurations. In addition, it is not uncommon for a UE to request an emergency VoLTE fallback through the \emph{Service Request for Emergency} and allow LTE to handle the voice call. For instance, Figure~\ref{fig:prackfallback} shows an SIP PRACK attempt by the UE after an EPS fallback due to our attack on 5G cells. However, even this mechanism can be impacted as the attacker can continue the DoS and potentially operate another false LTE cell for further exploitation. To further intensify the attacks, an adversary could also operate multiple rogue base stations supporting different generations (\eg, 4G, 3G and 2G) and multiple frequency bands. In case the UE attempts a fallback mechanism to previous radio access technologies, the adversary may still be able to attack the user. As a result, the user may not have access to any emergency features.

\begin{table}[tb!]
\centering
\caption{\label{tab:spoofing-conf} Used spoofing configurations and techniques. We classify them into sufficient and maximum impacts.}
\Large
\resizebox{\columnwidth}{!}{%
\begin{tabular}{@{}l|c|c}
\toprule
\textbf{PWS Spoofing conf. \& tech.} & Sufficient Impact & Maximum Impact \\ \hline \hline
\\[\dimexpr-\normalbaselineskip+2pt]
\textbf{SI Periodicity} & $16$ frames & $512$ frames \\
\\[\dimexpr-\normalbaselineskip+2pt]
\textbf{Repetition Period} & $10$ & $131,071$ \\ 
\\[\dimexpr-\normalbaselineskip+2pt]
\textbf{Number of Broadcasts} & $10,000$ times & $65,535$ times \\ 
\\[\dimexpr-\normalbaselineskip+2pt]
\textbf{Concurrent Warnings} & \shortstack{no} & \shortstack{yes} \\
\\[\dimexpr-\normalbaselineskip+2pt]
\textbf{Message ID Permutations} & \shortstack{no} & \shortstack{yes}  \\
\\[\dimexpr-\normalbaselineskip+2pt]
\textbf{Serial No. Permutations} & \shortstack{no} & \shortstack{yes} \\
\\[\dimexpr-\normalbaselineskip+2pt]
\textbf{Max Segment Length} & $32$ bytes & $32$ bytes \\
\bottomrule
\end{tabular}
}
\end{table}

%% file: sections/6.countermeasures.tex
\section{Countermeasures} \label{count-fbs}

We next discuss possible countermeasures aiming to detect or prevent the presented attacks. 

\noindent \textbf{Partial PKI-based Countermeasure.}
3GPP's study on 2G-4G~\cite{3gpp.133.969} is encouraging the adoption of a Public Key Infrastructure (PKI) for signing and verifying the SIB messages responsible for delivering alerts in HPLMN and VPLMN. The UE will be provided with a public key in order to validate the signed warning messages; the UE will need to be updated whenever the key or algorithm configurations change. SIB transmissions as illustrated in Figure~\ref{fig:warning_idle_inactive_connected} will be signed by the network's private key. 3GPP has proposed several techniques to address secure key provision on 2G, 3G and 4G (but not 5G), \ie, implicitly installed CA certificates on UE, over-the-air key distribution via Protocol Data Unit (APDU) commands~\cite{3gpp.131.115, 3gpp.131.116, 3gpp.131.102, 3gpp.133.969}, distribution through the General Bootstrapping Architecture (GBA)~\cite{3gpp.133.220, 3gpp.133.969},
and through NAS.

However, the implementation of such a system faces maintenance and operational hurdles. It requires adoption by all; HPLMN, VPLMN and UE. If the UE is designed to verify messages with other key and algorithm parameters than VPLMN's, the VPLMN public key is not available, there is no efficient way to distribute the public key to the UE, or the VPLMN does not support verification, then this will result in failures and broken security. Key distribution may encounter issues as well. For instance, an explicit TAU does not exist in 5G to be used for key delivery, and implicitly installed certificates from a Certificate Authority (CA) may induce issues with the sharing CAs among operators in various countries introducing new national threats.
Moreover, this mechanism may be inappropriate for security altogether. Since only SIB 6, 7, and 8 are protected, the attacker can still abuse the other broadcast messages (\eg, MIB and SIB 1) and further security flaws from Section~\ref{flaws} remain unmitigated. In fact, the barring attack and the malicious attachment persist with their associated impact. Spoofing can be avoided only if the UE is configured to deny any unauthenticated messages and the PLMN always signs the messages correctly. 

Table~\ref{tab:roaming} presents the effectiveness of this defensive mechanism while taking into account our attacks. This includes verification support by the network (signing the messages with the private key, first column in Table~\ref{tab:roaming}) and verification support by the UE (applying the network's public key to verify the messages, second column in Table~\ref{tab:roaming}). 
For each combination of the first two columns, Table~\ref{tab:roaming} specifies the feasibility of spoofing, suppression and rejection of legitimate messages which leads to user exposure. The first row portrays the current PWS implementation which is susceptible to spoofing and suppression, but false rejection is not possible since the UE accepts all messages even if the PLMN does not support PWS completely. When the UE does not support verification of the warning messages (\ie, rows 1 \& 3), spoofing is possible since verification never takes effect, allowing all messages. In contrast, spoofing is not feasible if the UE is strictly verifying all messages (\ie, rows 2 \& 4). However, when the PLMN does not support the verification scheme or there is no compatibility, false rejection of legitimate messages can occur (\ie, row 2). On top, suppression is not prevented, impacting verified and unverified warning messages.

\noindent \textbf{Full PKI-based Countermeasures.} Instead of protecting only warning-based SIB messages by a partial PKI-based countermeasure (with all the described disadvantages), a more viable solution may be full PKI-protection for all MIB and SIB messages, as also mentioned in~\cite{3gpp.33.809}. This will deprive the attacker the capability of imitating a legitimate base station from the beginning. However, the performance overhead for the certificates, distribution, maintenance, revocation, architectural redesigns, post-quantum solutions, and legacy device support have not been evaluated on real 5G networks to better comprehend this PKI's benefits and drawbacks.

On top of that, current optimised verification proposals for SIB 1 only~\cite{Ankush21:secure-bootstrapping, Hussain19:Insecure-Bootstrapping} are not adequate, as the PWS barring attack could still be feasible because of the exposed MIB. Additionally, the \texttt{cell\_barred} and \texttt{intra\_freq\_reselection} have moved from SIB 1 to MIB on 5G architecture, indicating the importance of a holistic defensive mechanism for MIBs and SIBs likewise.  

\noindent \textbf{Client-based countermeasures.} Client-based approaches and base station detection have been suggested~\cite{Lee19:spoofing-alerts-lte, Jihoon21:Securing-Wireless-Alerts, Li2017FBSRadarUF, 206168, 2102.08780, arif2020}. Such approaches rely on signature-based or Machine Learning (ML) models, but they may not be enough to ward off PWS attacks completely. First, emergency warning attacks are not entirely included in the detection mechanisms, as detection concerns mainly the presence of fake base stations and the state of insecure/secure NAS and RRC connections. As a result, accurate detection may not be possible, and the lack of preventive countermeasures can still expose UEs to warning suppression attacks. Second, these mechanisms usually take the form of mobile applications which also alert the user about potential attacks. They are typically restricted to Android and may require constant root access to cellular information, which may not be available or desirable by the user. Third, every network topology has its own properties and each mechanism must adapt its heuristics accordingly, especially in the case of ML subject to false positives and negatives.

\begin{table}[tb]
\resizebox{\columnwidth}{!}{ 
\begin{tabular}{ccccc} 
\toprule 
\multicolumn{2}{c}{\textbf{PWS Defensive Measure}} & \multicolumn{3}{c}{\textbf{Attack Success}}\\
\cmidrule(lr){1-2}
\cmidrule(lr){3-5}
\textbf{Security } & \textbf{Signature } &  \textbf{Spoo-}  &  \textbf{Sup-} & \textbf{False }\\ 
\textbf{Support} & \textbf{Verification} & \textbf{fing} & \textbf{pression} & \textbf{Rejection}\\ 
\midrule

%\rowcolor{grey}
\xmark & \xmark &  \textcolor{red}{Yes} & \textcolor{red}{Yes} & No \\

\xmark & \cmark & No & \textcolor{red}{Yes} & \textcolor{red}{Yes} \\  

\cmark & \xmark &  \textcolor{red}{Yes} & \textcolor{red}{Yes} & No \\ 

\cmark & \cmark & No & \textcolor{red}{Yes} & No \\ \bottomrule

\end{tabular} 
}
\caption{Security results for PWS verification (HPLMN-VPLMN). The first row represents the current implementation of PWS that has no security verification. In all cases, the UE needs to have the capability to process and display warning messages (USIM structure~\cite{3gpp.131.102}).} \label{tab:roaming}
\end{table}

\begin{comment}
\begin{table*}[!ht]
\centering
\Large
\resizebox{\textwidth}{!}{ 
\begin{tabular}{|c|c|c|c|c|c|} \hline 
\textbf{PWS Security Support} & \textbf{PWS Signature Verification} &  \textbf{PWS Message Processing} & \textbf{Spoofing} & \textbf{Suppression} & \textbf{False Rejection}\\ \hline \hline

\xmark & \xmark & \cmark & \textcolor{red}{Yes} & \textcolor{red}{Yes} & No \\ \hline

\xmark & \cmark & \cmark & No & \textcolor{red}{Yes} & \textcolor{red}{Yes} \\ \hline 

\cmark & \xmark & \cmark & \textcolor{red}{Yes} & \textcolor{red}{Yes} & No \\ \hline

\cmark & \cmark & \cmark & No & \textcolor{red}{Yes} & No \\ \hline \hline

\end{tabular} 
}
\caption{Security Results for roaming and PWS verification. The first row represents the current implementation of PWS that has no security verification. The analysis applies to both HPLMN and VPLMN cases.} \label{tab:roaming}
\end{table*}
\end{comment}

\noindent \textbf{Full RRC/NAS protection.} Another preventive approach is the adoption of mandatory encryption and integrity-protection for all messages, in particular the unprotected RRC and NAS messages (also mentioned in Sec.~\ref{flaws}) in the control-plane traffic. Such an implementation prevents message manipulations and eliminates malicious attachments. However, SigOver~\cite{Yang19:signal-overshadowing} and SigUnder~\cite{LudantN21} techniques could still impact the network as they do not require UE attachment. Past literature has repeatedly proposed RRC and NAS protection, experimenting on LTE~\cite{Kim19:touching-untouchables, Hussain18:LTE-Inspector, Hussain19:5G-reasoner, Park22:Negative-test, Park22:Negative-test}, but 5G specification and implementations do not meet such requirements.

\noindent \textbf{Monitoring and Attack Detection.} One orthogonal approach to preventive measures is via measurement collection, reporting and monitoring. Enriched measurement reports~\cite{Bitsikas21:handover-exploitation, 3gpp.33.809} with extra security fields (\eg, MIB/SIB hashes or locations of base stations) could be as suitable candidate.

In the case of PWS, UEs having received warning messages could send hash digests of the received messages back to the core network via enriched measurement reports that aggregates them. Even if only some of the UEs would support such a functionality, the network could verify the legitimacy of alerts and make users aware of fake messages. Operators could also operate a public web page which users could use to cross-check the legitimacy of warning alerts, a short url link could be part of all legitimate warning messages. Authorities could be informed too about attacking incidents along with the cell locations included in the measurement reports.

%% file: sections/8.related-work.tex
\section{Related Work}

\noindent\textbf{Security of Broadcast and Paging Messages.} One of the earliest indications of broadcast security flaws and paging were investigated by Hussain~\etal~\cite{Hussain18:LTE-Inspector, Hussain19:5G-reasoner}, however the studies mainly focused on LTE and there was no exploration of PWS exploitation. The SigOver attack~\cite{Yang19:signal-overshadowing} focuses on physical-layer overshadowing, which allows an adversary to abuse SIB and paging messages on LTE by injecting a crafted subframe that exactly overshadows the legitimate one. This approach can be efficient due to its low requirements (\ie, low power consumption, unaffected by UE states, and low setup complexity) and stealthiness. In our case, we were able to achieve 100\% success rate for the PWS barring attack with just $10dB$ and $30dB$ for spoofing, which is less than the $40dB$ requirement specified by SigOver, while maximizing the spoofing capacity (Table~\ref{tab:spoofing-conf}) and duration. In addition,~\cite{LudantN21} proposes the SigUnder attack performing significant improvements on physical-layer overshadowing attacks which are capable of disallowing cell access and reselection. With proper adaptations, we believe that such techniques could be used against the PWS as well. Susceptibility of the paging messages in general has also been demonstrated in terms of privacy and DoS~\cite{Kaiming20:paging-storm-attack, Shaik16:practical-attacks-lte, Hussain19:privacy-attacks-paging}. On the defense side, Ericsson's study on paging~\cite{EricssonReport}, and paging protections~\cite{Ankush20:protect-paging} by Ankush \etal have proposed countermeasures attempting to hinder paging attacks.

\noindent\textbf{Security of the Emergency Systems.} 3GPP~\cite{3gpp.133.969} maintains a conceptual study on PWS where security deficiencies and suggested countermeasures are discussed. Nevertheless, this study is limited in terms of experimentation, accurate attack definition, evaluated impact, and lacks 5G security assessment. Furthermore, Lee~\etal~\cite{Lee19:spoofing-alerts-lte} has provided notable results on CMAS spoofing and attacker's range on LTE but the investigation remains limited to specific cases, to one generation and to one attacker setup. As a consequence, an accurate presentation of all attacker's capabilities is missing, as in this work we have unearthed multiple attacks, network setup cases, and warning messages on 5G. Finally, work has been conducted to assess emergency call resilience against DoS/DDoS~\cite{Hou21:emergency-calls, Onofrei10:DDoS-IMS, Guri17:911-DDoS}.

\noindent\textbf{5G Security Studies.} The resilience of 5G AKA procedure was explored by Basin~\etal~\cite{Basin18:5G-Auth} and Borgaonkar~\etal~\cite{Borgaonkar19:5G-AKA}, revealing potential security defects. Bitsikas~\etal \cite{Bitsikas21:handover-exploitation} demonstrated the exploitation of the handover procedure on 5G and LTE allowing an attacker to perform a MitM or DoS attack. Chlosta~\etal~\cite{Chlosta21:SUCI-catchers} and Haque~\etal~\cite{Abida21:Device-Auth} exploited the Subscription Concealed Identifier (SUCI) identifier and Permanent Equipment Identifier (PEI), respectively. Security issues on 5G RRC and NAS messages were investigated~\cite{Hu19:5G-NAS, Hussain19:5G-reasoner} but actual experimentation is needed with a 5G SA setup to fully explore the security flaws.

\noindent\textbf{LTE Flaws and Misconfigurations.} Security in the control plane traffic has been explored~\cite{Shaik19:device-capabilities, Kim19:touching-untouchables, Hussain18:LTE-Inspector, Park22:Negative-test, Chen21:Doc-Analysis-LTE} revealing major vulnerabilities, while some remain unmitigated until the new 5G standards. Moreover, Rupprecht~\etal~\cite{rupprecht19:layer-two, rupprecht20:imp4gt} identified layer two vulnerabilities leading to user plane exploitation and MitM attacks, while network misconfigurations on LTE have been confirmed~\cite{Chlosta19:Misconf-LTE} showing that implementation is as important as the specifications.

%% file: sections/9.conclusion.tex
\section{Conclusion}

In this work, we explored the security of the 5G warning system. We have identified the underlying vulnerabilities revealing that the PWS is exposed to suppression and spoofing attacks with detrimental results to the safety of the users while deploying different attacker setups. Specifically, the PWS barring attack is a perilous threat to a cellular environment since it does not demand excessive skills, equipment capabilities and configurations. Furthermore, we assessed the impact of the aforementioned attacks in roaming cases and when PWS performs warning verification. Finally, we discussed several countermeasures that could be deployed to make the PWS more resilient against adversaries.

%% file: sections/appendix.tex
\section{False Base Station Setup} \label{setup}

First, the adversary will perform a comprehensive investigation of the operator and cellular network in order to collect sufficient intelligence about the possible target areas and their configurations. This is important since operators in various countries may configure the RAN and PWS differently. Specifically, for cellular configurations the attacker will require the locations of the gNodeBs, the Cell Identifier, Tracking Area Identifier (TAI) which incorporates the Mobile Country Code (MCC), Mobile Network Code (MNC) and Tracking Area Code (TAC), Absolute Radio Frequency Channel Number (ARFCN), PRACH Root Sequence Index, and the supported services for 5G. Additionally, it is important to capture the MIB and SIB messages of the gNodeBs in order to later replay them with a signal strength higher than the legitimate base station in order to attract the victim-UEs. Once collected, the attacker can decide which geographical area to impact and imitate the corresponding gNodeB in that area. Using real configurations is more advantageous for the attacker since invalid ones, such as wrong Cell Identifiers, may lead to easier detections and more network errors during a malicious handover or cell reselection. Therefore, the attacker needs to imitate the behavior of a legitimate station as closely as possible and respond to UEs in all the vital RRC and NAS procedures. If necessary, the attacker could also use more than one base station to achieve higher coverage.

Apart from the cellular configurations, the attacker will study the behavior of the PWS in that specific country. This includes the types of messages that are usually broadcasted, the periods of the year that normal emergencies/incidents occur, the most commonly impacted geographical locations, the warning message structure and configurations (\eg, broadcasted text and periodicity). Consequently, the attacker will be able to adapt appropriately and apply close-to realistic warning configurations to avoid trivial detection. The attacker may also deploy several stations and perform other supplementary attacks in conjunction with warning spoofing or suppression to bolster the attack’s efficiency (\eg force cell search) and affect more users.

\section{Adoption of the PWS} \label{pws-adopt}

Deployment of PWS systems is widely increasing since 2009~\cite{One2Many}. As per 11 Dec 2018~\cite{EU-Directive}, all EU member states are obliged to have a public warning system in place by 21 June 2022 (including the European Economic Area Agreement countries) to protect EU citizens. At least one form of PWS has already been implemented in the US (CMAS), Canada (WPAS), Chile and Peru (LAT-ALERT), UAE (UAE-Alert), China, South Korea (KPAS), India (ITEWC), Japan, Singapore, Saudi Arabia, Oman, the Philippines, Indonesia, Sri Lanka, New Zealand, Australia, Taiwan, Mexico, Russia, and Turkey, while others are planning to implement and activate one soon (\eg, United Kingdom).
In general, countries that are susceptible to extreme events (\eg, weather and climate) are strongly in favor of such a system for public safety~\cite{everbridge}. Furthermore, the United Nations encourages adoption of warning systems due to the influx of climate events~\cite{united-nations, meteo}.

\begin{figure}[!ht]
    \centering
    \subfloat[\centering CMAS message]{{\includegraphics[width=.3\columnwidth,keepaspectratio]{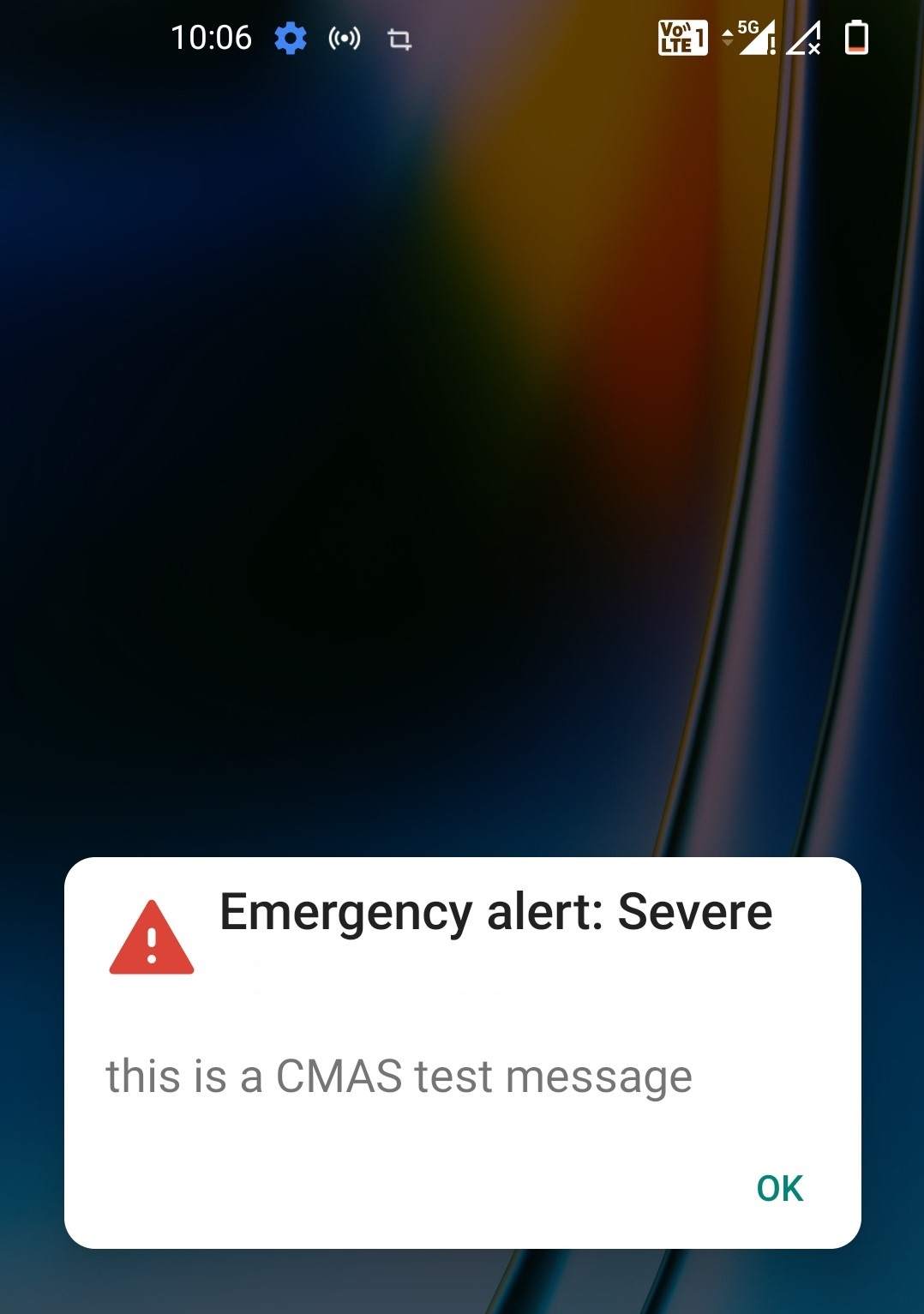} }}%
    \qquad
    \subfloat[\centering ETWS message]{{\includegraphics[width=.3\columnwidth,keepaspectratio]{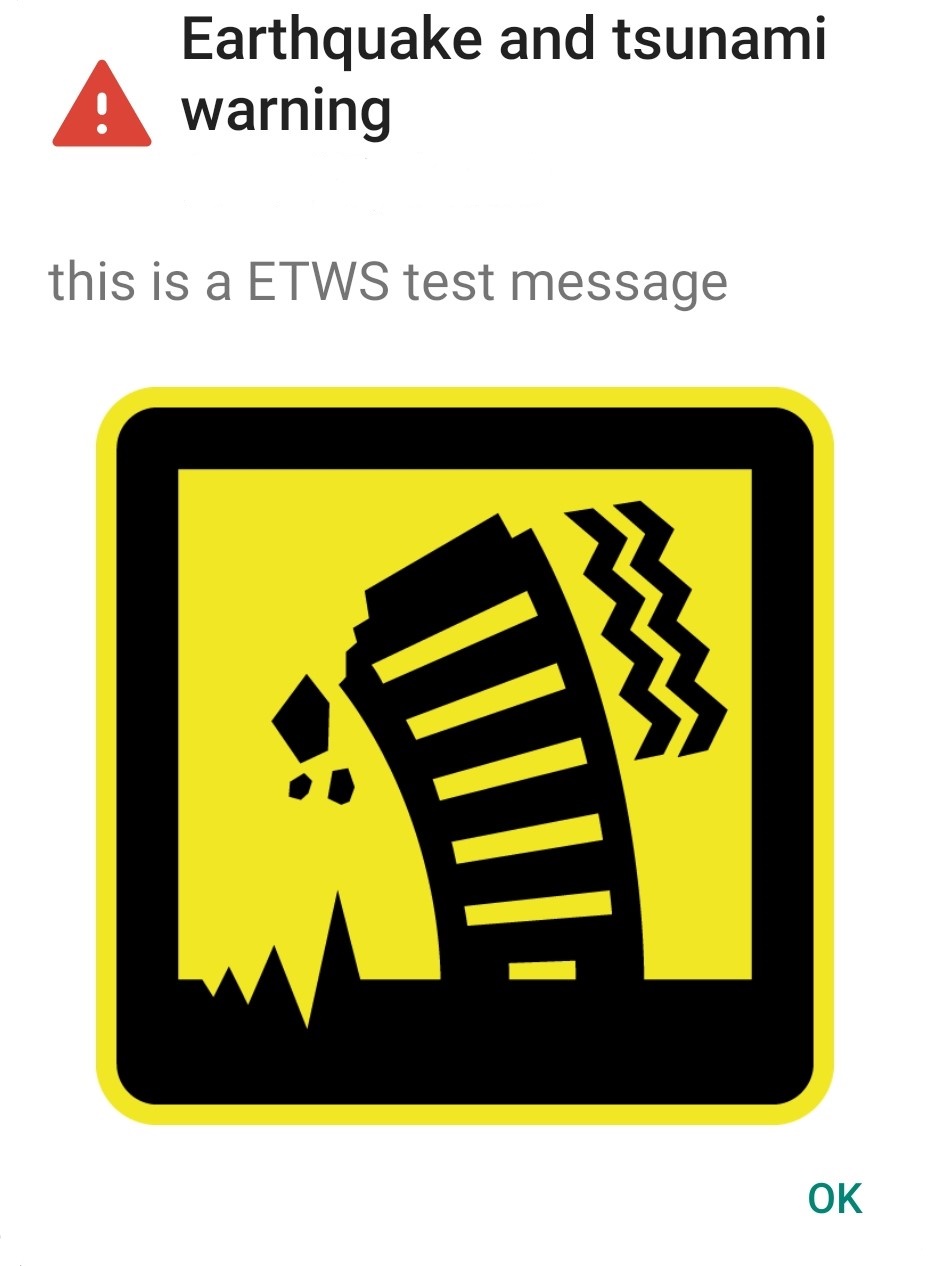} }}%
    \caption{Warnings on OnePlus Nord 2 5G}
    \label{fig:alerts-examples}%
\end{figure}

\section{Further Discussion} \label{further-discussion}

\textit{Is it a relevant threat if an attacker sends forged warning messages to an individual user or a small group of users, or only to a large crowd?} We believe that the more users receive fake messages the merrier the impact becomes as mass panic may lead to hazardous incidents such as physical harm. In fact, the attacker's objective is to affect as many people as possible within a dense, limited or overcrowded space for a specific time window. Geographically speaking,~\cite{Lee19:spoofing-alerts-lte} has shown that four LTE base stations can easily impact a stadium with an approximate capacity of $50,000$ seats. On such a level, we believe that the attacks can turn into an alarming risk.

\textit{Is it a relevant threat if an attacker can suppress warning messages to an individual user or a small group of users, or only to a large crowd?} Similar to spoofing bogus messages, message suppression aims to influence as many network subscribers as possible. Not allowing many users to get notifications about an emergency could also lead to life threatening scenarios. In addition, based on Section~\ref{exploitation}, we consider suppression more trivial than any spoofing approaches, hence increasing the threat level. 

\textit{Could PWS Security be a mandatory requirement?} Protection of warning notifications and users against attackers are subject to regulatory policies. According to 3GPP~\cite{3gpp.133.969}, the requirements for PWS security are optional since there are regions and countries that do not require or aim to deploy this feature any time soon (such as US and Japan). Therefore, without outright and compulsory security, networks will remain exposed to warning spoofing and suppression.

\textit{Are SMS-based warnings affected?} On 5G, the SMS~\cite{3gpp.123.040} deliveries are either over IMS (SIP-based) or over NAS. The UE receives a specific paging message for an SMS service and then the UE enters the RRC-Connected state while sending a Service Request. Then, the network delivers the SMS notifications using the downlink transmission. We ran dedicated experiments for the SMS-based warnings. Since our suppression attacks disallow connection to the legitimate network, the victim-UE will not be able to receive paging messages and consequently the SMS-based warning. Regarding spoofing SMS warnings and DoS, past works~\cite{Hua16:IMS-threats, Hongil15:Breaking-voLTE, Hua16:IMS-SMS, Traynor12} have demonstrated the weakness of SMS over IMS on LTE which may also affect 5G in certain cases. On the other hand, according to 3GPP~\cite{3gpp.33.501}, SMS over NAS implements encryption and integrity-protection when the UE has already activated NAS security with the AMF. However, more testing is required to verify its robustness on 5G against spoofing as~\cite{Kim19:touching-untouchables, Park22:Negative-test} have confirmed SMS over NAS weaknesses on LTE.

\section{Emergency Call Flow} \label{Control-flow}

\begin{figure}[t]
    \centering
    \includegraphics[width=.8\columnwidth,keepaspectratio]{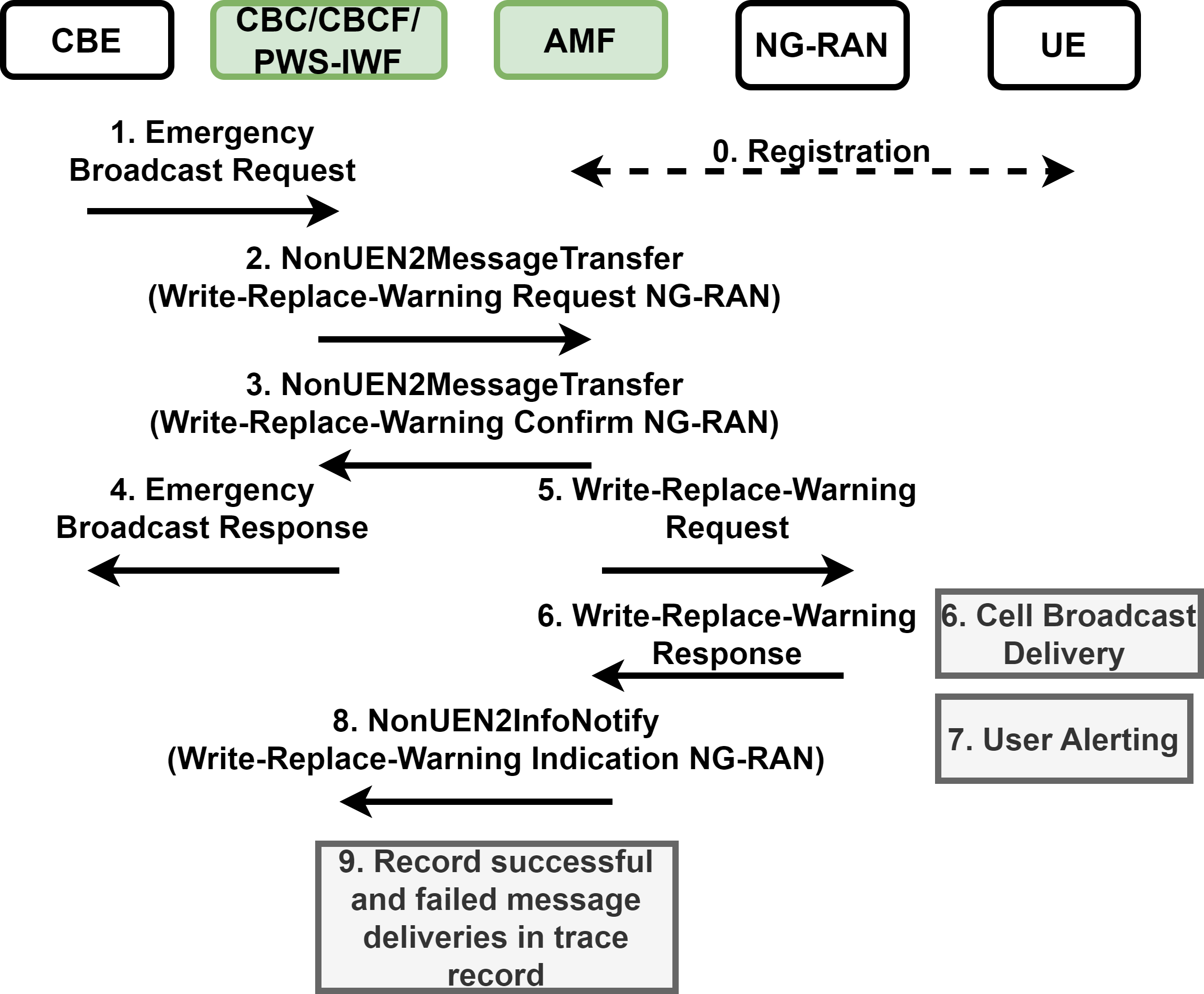}
    \caption{Emergency Call Flow}
    \label{fig:pwsgraph}
\end{figure}

The emergency warning procedure of Figure~\ref{fig:warning_idle_inactive_connected} is extended to Figure~\ref{fig:pwsgraph} having the following steps: 

\begin{enumerate}
\setcounter{enumi}{-1}

\item The UE registration and mutual authentication procedures are performed. Encryption and integrity protection are enabled for the established communication (Control plane and User plane) based on the specifications~\cite{3gpp.33.501}. 

\item The CBE sends the Emergency Broadcast Request to the CBC/CBCF based on the authorities. Then, the CBC/CBCF authenticates this request, which includes the warning type, warning message, impacted area and time period.

\item Using the area of impact, the CBC/CBCF identifies which AMFs need to be contacted and determines the information to be incorporated into the Warning Area List NG-RAN Information Element (IE). The CBC/CBCF sends a Write-Replace-Warning Request message containing the warning message and its attributes to the AMFs. In case of a PWS-IWF entity, the message is forwarded through it to the AMFs. Additionally, the Write-Replace-Warning Request message usually incorporates the Message Identifier, Serial Number, list of NG-RAN Tracking Areas, Warning Area List NG-RAN, Global RAN Node ID, Warning Area Coordinates, etc.

\item The AMF sends a Write-Replace-Warning Confirm NG-RAN message indicating to the CBC/CBCF that the AMF has started the distribution of warning message to NG-RAN base stations. The Write-Replace-Warning Confirm NG-RAN message may contain the Unknown Tracking Area List IE, which identifies the Tracking Areas that are unknown to the AMF and where the Request cannot be delivered. If this message is not received by the CBC/CBCF within an appropriate time period, the CBC/CBCF may attempt to deliver the warning message via another AMF in the same region.

\item Upon reception of the Write-Replace Confirm NG-RAN messages from the AMFs, the CBCF may confirm to the CBE that the distribution of the warning message has commenced.

\item The AMF forwards Write-Replace-Warning Message Request NG-RAN to NG-RAN nodes. If the list of NG-RAN Tracking Areas is not included and no Global RAN Node ID has been received from the CBC/CBCF, the message is forwarded to all RAN base stations that are operated under the AMF. On the other hand, if a Global RAN Node ID has been received, the AMF shall forward the message only to the NG-RAN base station indicated by this ID IE.

\item The NG-RAN node first detects duplicate messages by checking the message identifier and serial number fields within the warning message as it may receive the same message from multiple AMFs. If any redundant messages are detected, only the first one received will be broadcasted by the cells. The NG-RAN base station shall use the Warning Area List NG-RAN information to determine the cell(s) in which the message is to be broadcasted. Furthermore, the NG-RAN base stations return a Write-Replace-Warning Message Response to the AMF, even in case of a duplicate. If there is a warning broadcast message already ongoing and the Concurrent Warning Message (CWM) Indicator is included in the Write-Replace-Warning Request NG-RAN message, the base station does not stop the existing broadcast message but starts broadcasting the new message concurrently. Otherwise, it shall immediately replace the existing broadcast message with the newer one. If concurrent warning messages are not supported, message priority is enforced. Eventually, each base station begins delivering the paging and the SIB messages to all available UEs as illustrated also by Figure~\ref{fig:warning_idle_inactive_connected}.

\item If the UE has been configured to receive warning messages, and to accept warnings on that PLMN~\cite{3gpp.131.102}, then the UE can use warning type values, such as 'earthquake' or 'tsunami', immediately to alert the user. When the warning type is 'test', the UE silently discards the primary notification, but the specially designed UEs for testing purposes may alert the user based on his/her coordinates and location. At the same time, the Write-Replace Response is sent to the AMF as an acknowledgement.

\item If the Warning-Message-Indication parameter was present in the Write-Replace-Warning Request NG-RAN and it is configured in the AMF based on operator's policy, the AMF shall forward the Broadcast Scheduled Area Lists in a Write-Replace-Warning Indication(s) NG-RAN to the CBC/CBCF. The Broadcast Scheduled Area List shall contain the Broadcast Completed Area List the AMF has received from the NG-RAN node. Nevertheless, this step is optional.

\item From the Write-Replace-Warning Response messages returned by NG-RAN base stations the AMF determines the success or failure of the delivery and creates a trace record. 
\end{enumerate}

\section{Network Configurations} \label{network_conf}

We configured the network to use the testing PLMN which is $00101$ and to consist of a gNodeB and a Core Network (including AMF and IMS) in the Amarisoft Box. The gNodeB had the following configurations: $gnb\_id = 0x1234A$, $tac = 100$ (decimal), and $root\_sequence\_index = 1$. It also included two distinct cells; cell $1$ with $cell\_id = 0x01$ and $n\_id\_cell = 500$, and cell $2$ with $cell\_id = 0x02$ and $n\_id\_cell = 501$. Both cells possessed separate frequencies in the $n78$ band for 5G Standalone and in the $n41$ band for 5G non-Standalone. During our experimentation we tested both the Time Division Duplex (TDD) and Frequency Division Duplex (FDD) communication technologies. For testing emergency SMSs and SIP calls we also incorporated an IP Multimedia System (IMS) into the core network which allowed us to send SMSs to the UE and allowed the user to call $911$, as it is supported by Amarisoft. Finally, for the UEs we had to configure the Access Point Names (APNs) in order for the UEs to be fully connected. According to Amarisoft documentation we used the Internet APN and the IMS APN, whenever possible.

\section{CMAS and ETWS Configurations}\label{warning_conf}

The basic format of our ETWS/CMAS messages follows the Amarisoft guidelines. During our experimentation we altered only values that are presented in the following structures in respect to the specifications:

pws\_msgs: [ 
\tabto{38pt}     \{ \textbf{/* ETWS earthquake or tsunami message */} \\
\tabto{38pt}    local\_identifier: 1, \\
\tabto{38pt}    message\_identifier: 0x1102, \\
\tabto{38pt}    serial\_number: 0x3000, \\
\tabto{38pt}    warning\_type: 0x0580,  \\
\tabto{38pt}    data\_coding\_scheme: 0x0f, (GSM 7 bit encoding) \\
\tabto{38pt}    warning\_message: "This is a ETWS test message" \},\\
\tabto{38pt}    \{ \textbf{/* CMAS Presidential Level Alert */} \\
\tabto{38pt}    local\_identifier: 2, \\
\tabto{38pt}    message\_identifier: 0x1112, \\
\tabto{38pt}    serial\_number: 0x3000, \\
\tabto{38pt}    data\_coding\_scheme: 0x0f, (GSM 7 bit encoding) \\
\tabto{38pt}    warning\_message: "This is a CMAS test message"\} ]  \\

\section{Experimental Evidence}

\begin{figure}[!hbt]%
    \centering
    \subfloat[\centering CMAS message]{{\includegraphics[width=.4\columnwidth,keepaspectratio]{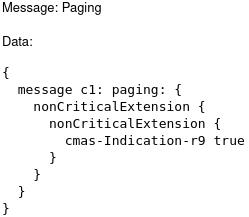} }}%
    \qquad
    \subfloat[\centering ETWS message]{{\includegraphics[width=.4\columnwidth,keepaspectratio]{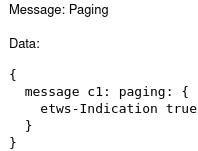} }}%
    \caption{Paging Messages}
    \label{fig:paging}%
\end{figure}

\begin{figure}[htb]
    \centering
    \includegraphics[width=0.45\columnwidth,keepaspectratio]{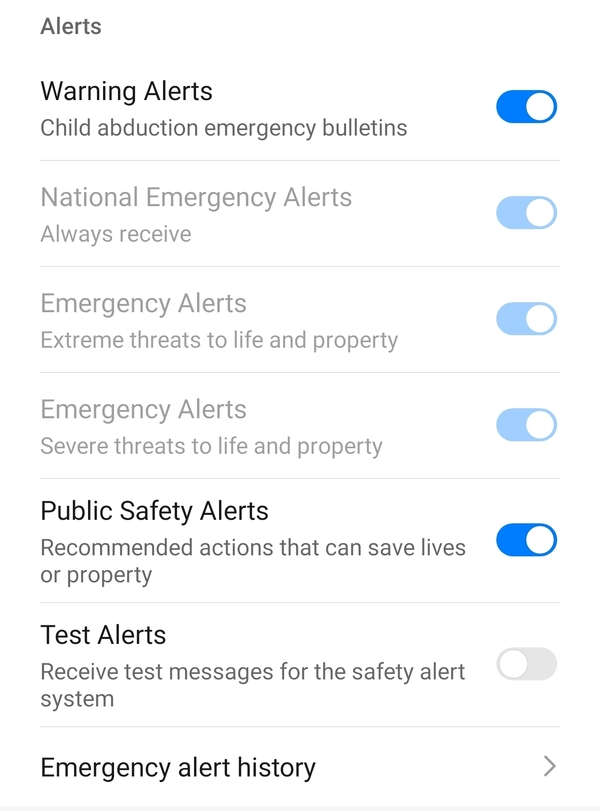}
    \caption{Compulsory \& Optional Alert on Huawei P40 5G}
    \label{fig:alert-options}
\end{figure}

\begin{figure}[!hbt]
    \centering
    \includegraphics[width=.9\columnwidth,keepaspectratio]{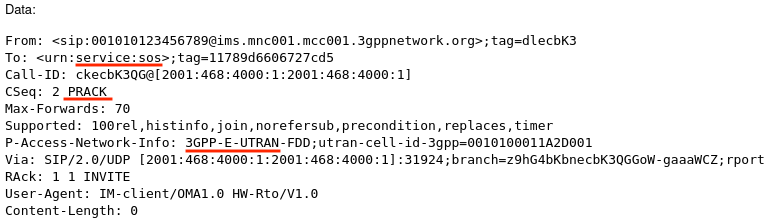}
    \caption{SIP PRACK message for an emergency call and VoLTE Fallback.}
    \label{fig:prackfallback}
\end{figure}

\begin{figure}[!hbt]
    \centering
    \includegraphics[width=.8\columnwidth,keepaspectratio]{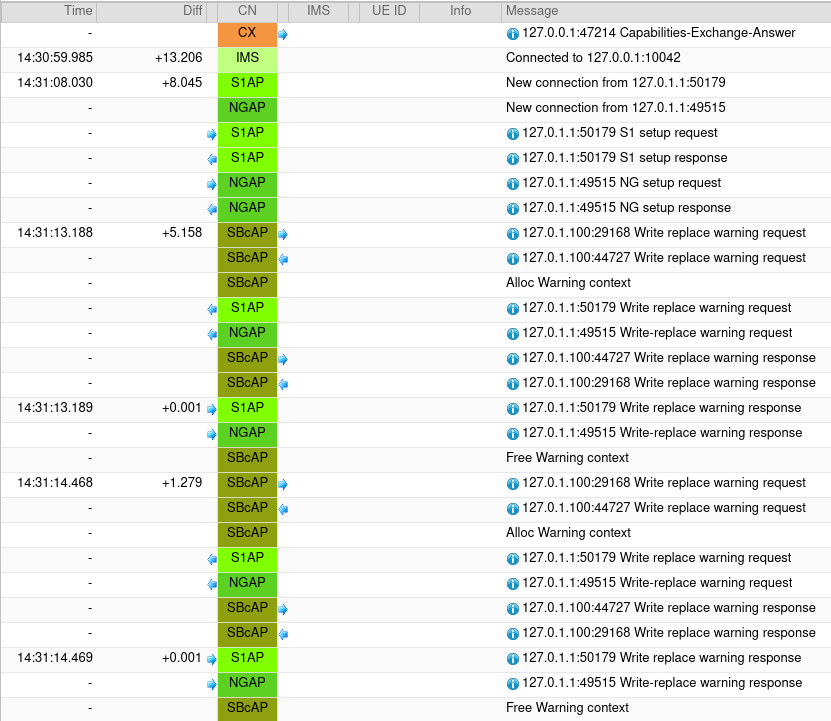}
    \caption{Warning Flow for the CBC, Core Network and RAN.}
    \label{fig:warningflow}
\end{figure}

\begin{figure}[!hbt]
\minipage{.3\textwidth}
  \includegraphics[width=\linewidth]{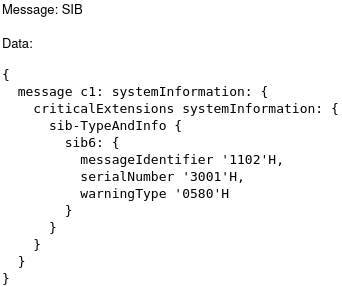}
  \caption{SIB6}\label{fig:sib6}
\endminipage\hfill
\minipage{.45\textwidth}
  \includegraphics[width=\linewidth]{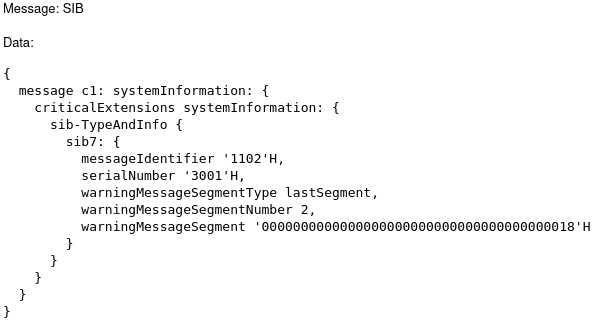}
  \caption{SIB 7}\label{fig:sib7}
\endminipage\hfill
\minipage{.45\textwidth}%
  \includegraphics[width=\linewidth]{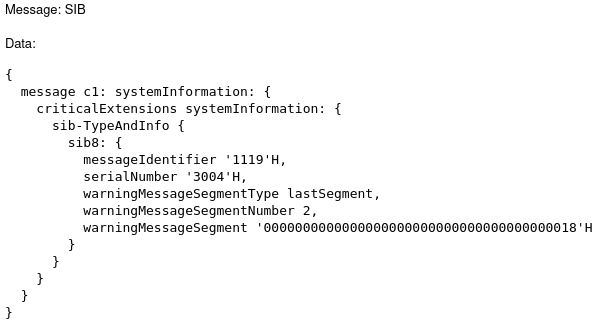}
  \caption{SIB 8}\label{fig:sib8}
\endminipage
\end{figure}

\begin{figure}[!hbt]
    \centering
    \includegraphics[width=0.9\columnwidth,keepaspectratio]{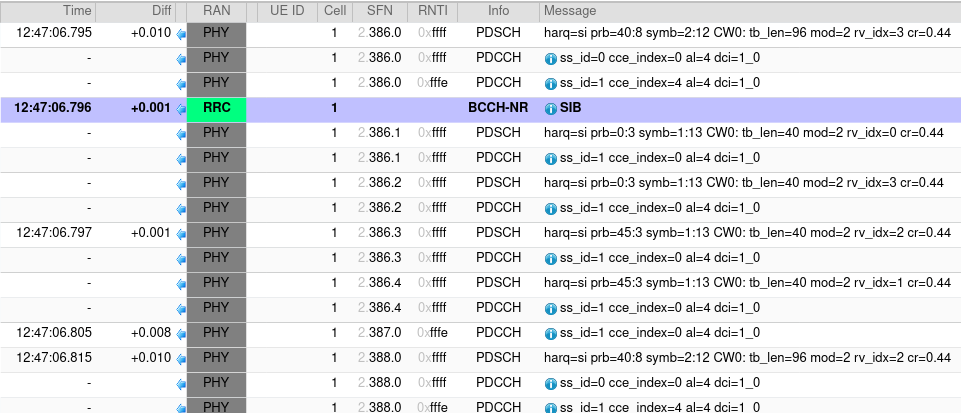}
    \caption{Broadcast Control Channel (BCCH) used for SIB 6 transmission.}
    \label{fig:sibflow}
\end{figure}

Figures~\ref{fig:sibflow} and~\ref{fig:warningflow} show the transmitted warnings in our setup with all the involved network entities.  Figures~\ref{fig:paging} and ~\ref{fig:sib6}-\ref{fig:sib8} show the structure of the paging and SIB messages that we used as a core network and as an attacker. Figure~\ref{fig:prackfallback} presents a SIP emergency call during our attacks on 5G PWS. Finally, Figure~\ref{fig:alert-options} displays the discretionary and mandatory options on Huawei P40 5G.

